\documentclass[12pt]{article}
\pdfoutput=1
\usepackage{amssymb,amsmath,amsfonts}
\usepackage{graphicx}
\usepackage{color}
\usepackage{enumitem}
\usepackage[colorlinks=true
,urlcolor=blue
,citecolor=blue
,linkcolor=blue
,pagecolor=blue
,linktocpage=true
,pdfproducer=medialab
]{hyperref}
\usepackage[a4paper]{geometry}
\usepackage{enumitem}

\usepackage{cite}
\usepackage[section]{placeins}
\usepackage{slashed}
\usepackage{mathtools}
\usepackage[FIGTOPCAP]{subfigure}

\makeatletter \renewcommand{\@dotsep}{10000} \makeatother

\mathchardef\mhyphen="2D

\def\422{$4\mhyphen2\mhyphen2$}
\newcommand{\beq}{\begin{equation}}
\newcommand{\eeq}{\end{equation}}
\newcommand{\bea}{\begin{eqnarray}}
\newcommand{\eea}{\end{eqnarray}}

\usepackage[nodisplayskipstretch]{setspace}

\begin{document}

\begin{titlepage}
\pagestyle{empty}

\vspace*{0.2in}
\begin{center}
{\Large \bf Scalar Dark Matter and Electroweak Stability}\\
\vspace{1cm}
{\bf Durmu\c{s} Demir$^{a,}$\footnote{Email: durmus.demir@sabanciuniv.edu}} 
and
{\bf Cem Salih {\"U}n$^{b,}$\footnote{E-mail: cemsalihun@uludag.edu.tr}}
\vspace{0.5cm}

{\it $^a$Faculty of Engineering and Natural Sciences,
  Sabanc{\i} University,
  34956  Tuzla, {\.I}stanbul, Turkey}, \\

{\it $^b$Department of Physics, Bursa Uluda\~{g} University, TR16059 Nil\"{u}fer,  Bursa, Turkey.}

\end{center}

\vspace{0.5cm}
\begin{abstract}
\noindent
The standard model of elementary particles (SM), despite experimental completion at the LHC, needs to be extended for various physical reasons, including the cold dark matter (DM). Each extension comes with its scale and mechanism, and typically lifts, at the loop level, the electroweak scale towards its high scale. The problem is to keep the electroweak scale stable while providing a room for the aforementioned heavy extensions. To this end, it turns out that the SM Higgs sector remains stable in the presence of a heavy scalar if their quartic couplings unify at a certain scale when their masses are degenerate. Under this mass-degeneracy-driven unification (MDDU),  the scalar under concern is found to qualify as a viable DM candidate and to leave the electroweak scale stable. Our detailed simulation studies explicitly show that the MDDU parameter space agrees with current collider and astrophysical bounds. Our work can be extended to other relevant scalars (like flavons, inflaton and others) as a mechanism by which the electroweak scale is held stable. 

\end{abstract}

\end{titlepage}

\setcounter{footnote}{0}
%%%%%%%%%%%%%%%%%%%%%%%%%%s
% Main body
%%%%%%%%%%%%%%%%%%%%%%%%%%

\section{Introduction}
\label{ch:introduction}

Even though the SM is experimentally completed after the discovery of the Higgs boson \cite{higgs}, it is far from being complete conceptually and theoretically as it suffers from various problems such as destabilizing UV sensitivities \cite{veltman}, exclusion of gravity \cite{grav}, and absence of a  suitable dark matter candidate \cite{dm-ilk}. The UV sensitivity of the Higgs boson mass is so strong that the SM is effectively forced to end at $\mathcal{O}(1~{\rm TeV})$ to get merged with physics beyond the SM (BSM) \cite{veltman}. However, the current LHC results have already shown that the SM continues to hold good up to energies well above the TeV border \cite{exotica}. This experimental fact sidelines known completions of the SM (supersymmetry, extra dimensions, compositeness, and their hybrids) as they can be relevant only if they lie at the scales of order a TeV. The reason is that their BSM sectors (superpartners in supersymmetry, Kaluza-Klein modes in extra dimensions, and technifermions in compositeness)  couple to the SM with SM-sized couplings ($\lambda_{SM-BSM}\simeq \lambda_{SM}$) so that heavier the BSM is larger the shift in the Higgs boson mass.  Thus, according to the LHC results \cite{exotica}, the SM can have heavy BSM completions
\begin{equation}
%\setstretch{2.5}
\begin{array}{ll}
(C_1) & {\rm if\ the\ destabilizing\ UV\ sensitivities\ of\ the\ SM\ are\ {\it transmuted,}\ and} \\\\
(C_2) & {\rm if\ the\ transmutation\ allows\ for\ sufficiently  {\it small}\ SM-BSM\ couplings~}.
\end{array}
\label{conditions}
\end{equation}
Here, the condition $(C_1)$ is satisfied ordinarily in all the known completions. The condition $(C_2)$, however, is highly nontrivial in that the known SM completions necessarily require $\lambda_{SM-BSM}\simeq \lambda_{SM}$, and never work in the small coupling regime, $\lambda_{SM-BSM}\ll \lambda_{SM}$. In fact, this hierarchic regime is what lies at the heart of the electroweak hierarchy problem \cite{veltman}. To reiterate, the extension of the SM that transmutes the UV sensitivities of the SM must work with and allow for hierarchically small SM-BSM couplings. This constraint and the  conditions  $(C_1)$ and $(C_2)$ in (\ref{conditions}), are satisfied presently by one UV completion: The symmergent gravity \cite{demir1,demir2}. In this particular completion,
\begin{equation}
%\setstretch{2.5}
\begin{array}{ll}
(P_1) & {\rm curvature\ emerges\ in\ a\ way\ restoring\ gauge\ symmetries,}\\ { }&{\rm transmutes\ destabilizing\ UV\ sensitivities\ into\ curvature\ terms,}\\ { }&{\rm and\ incorporates\ this\ way\  the\ GR\ into\ the\ SM\ such\ that}\\\\
(P_2) & {\rm BSM\ sector\ arises\ as\ a\ necessity\ with\ no\ need\ to\ any\ SM-BSM\ coupling.}
\end{array}
\label{properties}
\end{equation}
In $(P_1)$, gravity is incorporated into SM+BSM by taking a flat spacetime effective SM+BSM into a curved spacetime (not the SM+BSM itself, as in  \cite{grav}). In $(P_2)$, BSM is a renormalizable QFT of various massless, massive and ultra massive non-SM fields, where none of them has to interact with the SM fields for symmergence to work. They are thus largely unconstrained except that the mass matrix ${\mathcal{M}}$  of the SM + BSM fields must satisfy the sum rule ${\rm str}\left[{\mathcal M}^2\right]= 64 \pi^2 M_{Pl}^2$. This sum rule requires the BSM to be boson dominated either in mass or in number. Among all the BSM fields, scalars occupy a special place as they are necessary for realizing physical phenomena like mass generation, cosmic inflation, quintessence, and dark matter. And the scalar fields tend to destabilize the SM for the reasons stated in (\ref{conditions}).

It is with mechanisms like symmergence that SM-BSM couplings are loosened as in (\ref{properties}) to allow for a possible resolution of the hierarchy problem. In Sec. \ref{sec:linkup} below, for definiteness, we focus on a BSM scalar $S$ (as the only BSM field that couples to the SM), and analyze under what conditions it does not destabilize the SM Higgs sector. 

In general, as mentioned above, variety of phenomena can be studied in connection with scalar fields. We, nevertheless, specialize in Sec. 3 to cold dark matter (DM) \cite{dm-ilk} which,   as reviewed recently in \cite{dm-review},  can be modeled by a long-lived, neutral scalar $S$ of mass $m_S$ \cite{dm00,dm01,demir0,dm0}. Its longevity is ensured by its oddness under a $Z_2$ parity, which forbids its  decay channels (such as $S \rightarrow$ SM SM), and nullifies its  vacuum expectation value, that is, $\langle S \rangle =0$. 

In Sec. \ref{sec:model}, building on a detailed phenomenological analysis, we discuss the Higgs boson profile in terms of its invisible decays and exemplify its decays into certain SM particles. In Sec. \ref{sec:model}, again, we determine in what parameter regions the scalar $S$ qualifies as a viable DM candidate.  In regions where $S$ reproduces the measured relic density, we study its co-annihilation rates as well as its scattering rates from the nuclei. Finally, In Sec. \ref{sec:conc}, we summarize and conclude.

\section{Mass-Degeneracy-Driven Unification}
\label{sec:linkup}
The problem with scalar field theories is that they are self-destructively sensitive to the UV domain \cite{veltman}. 
Even if all the quadratic ($\Lambda_\wp^2$) and quartic ($\Lambda_\wp^4$) sensitivities to the UV momentum cutoff $\Lambda_\wp$ are transmuted into curvature terms to lead to the GR \cite{demir2}, the remnant $\log\Lambda_\wp$ sensitivities, which can be translated into dimensional regularization via $\log \Lambda_\wp^2 = {1}/{\epsilon} -\gamma_E + 1 + \log 4\pi Q^2$, lead to the  ${\rm \overline{MS}}$ correction ($S$ is the aforementioned BSM scalar)
\begin{eqnarray}
\label{corr-1}
\delta m_H^2 \propto \lambda_{H S} m_S^2 \log \frac{m_S^2}{Q^2}
\end{eqnarray}
which shifts the Higgs condensation parameter $m_H^2$ in proportion to $m_S^2\log m_S^2$ if the two scalars are coupled as $\lambda_{H S} H^\dagger H S^2$. The logarithmic corrections shown in (\ref{corr-1}) tend to displace the Higgs boson from its natural scale $m_H$ towards $m_S$ since larger the $m_S$  is larger the shift in $m_H$. There arises, of course, no problem when $m_S \sim m_H$. However, given the fact that the LHC have found so far no new particle \cite{no-S} beyond the SM, the DM particle is expected to be either heavy ($m_S\gg m_H$) or light ($m_S\ll m_H$) and, in both cases, there arises a severe hierarchy problem in the lighter scalar. This means that scale-split multi-scalar theories get degenerated by quantum corrections, and the resulting  mass instability in light scalars obstructs extension of the SM by new heavy scalars. 

Preventing this mass instability is a profound question, and its answer is both obvious and obscure. It is obvious in that $|\lambda_{HS}|$ must be just ``small" to start with since corrections to $\lambda_{HS}$ are proportional to $\lambda_{HS}$ itself. It is obscure in that there is no obvious selection rule or symmetry that can ensure the requisite ``smallness". Its dimensionless nature prevents also dynamical mechanisms like Giudice-Masiero mechanism \cite{g-m} because a promotion like $\lambda_{HS} \rightarrow H^\dagger H/S^2$ would simply mean  killing the coupling between $H$ and $S$. In this arid climate, the most one can do is to impose a judicious relationship among model parameters and try to secure it against renormalization group flow.  To this end, symmergence, which sets $\lambda_{HS}$ free (as opposed to the known completions which require $\lambda_{HS}$ to remain close to the Higgs quartic coupling $\lambda_H$), provides an eligible framework in which  $\lambda_{HS}$ can be linked to other model parameters to keep $\delta m_H^2$ under control. In this respect, the seesawic couplings introduced in \cite{demir1,demir2}
\begin{eqnarray}
\label{seesawic}
\lambda_{HS} \propto \lambda_H \frac{m_H^2}{m_S^2} 
\end{eqnarray}
possess  right structure to ensure that $\delta m_H^2$ in (\ref{corr-1}) remains below $m_H^2$. This seesawic relation, according to which heavier the $S$ smaller its coupling to $H$, arises as an additional constraint compared to the usual conditions
\begin{eqnarray}
\label{conds}
\lambda_H > 0\;,\;\; \lambda_H > 0\;,\;\;  16 \lambda_H \lambda_S - \lambda_{HS}^2 > 0
\end{eqnarray}
which ensure that the potential energy density
\begin{equation}
V(H,S) = -m_H^{2}H^{\dagger}H + \lambda_{H}(H^{\dagger}H)^{2}+ m_{S}^{2}S^{2}+ \lambda_{S}S^{4}+\lambda_{HS}H^{\dagger}HS^{2}
\label{eq:scalarpotential}
\end{equation}
is bounded from below to yield the Higgs boson $h=H_0-v_H$ of mass $m_h^2 = 2 m_H^2$ and the singlet boson $s=\sqrt{2}S$ of mass $m_s^2=m_S^2+ (\lambda_{HS}/2) \langle H_0 \rangle^2$ as two spinless  quanta excited from the vacuum configuration (with the Higgs VEV $\langle H_0 \rangle = m_H/\sqrt{\lambda_H}\approx 246\ {\rm GeV}$)
\begin{equation}
\label{vac}
\langle H\rangle = \frac{1}{\sqrt{2}}\left(\begin{array}{c} 0 \\ \langle H_0 \rangle \end{array} \right), \hspace{0.3cm} \langle S \rangle = 0 \hspace{0.5cm} 
\end{equation}
for which $V(H,S)$ is required to be invariant under $H\rightarrow H$ and $S\rightarrow -S$ (the $Z_2$ invariance with which the scalar $S$ will be taken as DM candidate in the next section).

The seesawic structure in (\ref{seesawic}) relates $\lambda_{HS}$ to the field masses. This means that 
for stabilizing the Higgs mass (suppressing $\delta m_H^2$ in (\ref{corr-1})) the parameters in the potential must exhibit a mass-dependent relationship beyond the energy conditions in (\ref{conds}). To this end, a {\it mass-degeneracy-driven unification} (MDDU) of the form (at a given scale $Q=Q_0$)
\begin{eqnarray}
\label{link}
\lim_{m_H(Q_0)\rightarrow m_S(Q_0)} \lambda_H(Q_0) = \lambda_S(Q_0) =|\lambda_{HS}(Q_0)|
\end{eqnarray}
proves to be a useful criterion as it possesses (among many) the particular solution
 \begin{eqnarray}
\lambda_S(Q_0) = \lambda_H(Q_0), \hspace{0.3cm}    |\lambda_{HS}(Q_0)| = \frac{2 \lambda_{H}(Q_0)}{\frac{m_H^2(Q_0)}{m_S^2(Q_0)} + \frac{m_S^2(Q_0)}{m_H^2(Q_0)}}
    \label{linkup}
    \end{eqnarray}
according to which $\lambda_{HS}(Q_0)$ reduces to the seesawic structure in (\ref{seesawic}) for  $m_S(Q_0)\gg m_H(Q_0)$, and smoothly covers the opposite limit of $m_S(Q_0)\ll m_H(Q_0)$. It obviously is not unique; one can consider numerous variations of (\ref{linkup}) that satisfy the MDDU in (\ref{link}). It is just one possible choice that keeps $\lambda_S$ and $|\lambda_{HS}|$ below the SM couplings. Its collider implications have already been investigated in  \cite{cankocak} at the seesawic limit (\ref{seesawic}). 

\begin{figure}[ht!]
\centering
\includegraphics[scale=0.8]{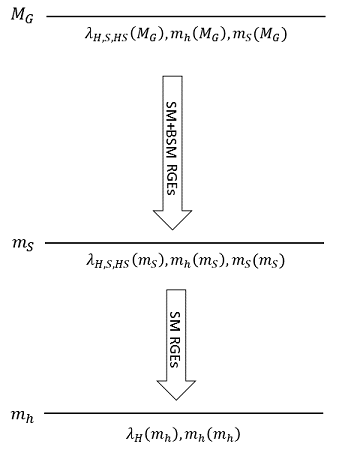}%
\caption{The RG evolution of the model parameters from $Q=Q_0=M_G$ down to $Q=m_h$. The scalar mass $m_S=m_S(m_S(M_G))$ is a divide between the SM and the SM+BSM.}
\label{fig:RGEevol1}
\end{figure}

The solution (\ref{linkup}) is a MDDU linkup at a scale $Q=Q_0$. The model parameters, the $m_H^2$ in particular, are computed at a given scale $Q$ via the renormalization group equations (RGEs) from $Q_0$ to $Q$. Hereon, we set  $Q_0=M_{G} \approx 10^{15}$ GeV (the GUT scale in non-SUSY models \cite{PerezLorenzana:1998rj}) in the MDDU in (\ref{linkup}). We give in Figure \ref{fig:RGEevol1} a schematic illustration of the renormalization group flow of the model parameters. With input data at $Q=Q_0\equiv M_G$, the parameters in the potential evolve from $Q=M_G$ down to $Q=m_S\equiv m_S(m_S(M_G))$ via the SM+BSM RGEs \cite{Haba:2016gqx}
\begin{eqnarray}
\label{RGE-lamH}
Q\dfrac{d\lambda_{H}}{dQ} &=& \left(Q\dfrac{d\lambda_{H}}{dQ}\right)_{SM} + \dfrac{1}{16\pi^{2}}\lambda_{HS}^{2}\\
\label{RGE-lamHS}
Q\dfrac{d\lambda_{HS}}{dQ} &=& \dfrac{\lambda_{HS}}{16\pi^{2}}\left(6\lambda_{H}-\dfrac{9}{2}g_{2}^{2}-\dfrac{3}{2}g_{1}^{2}+6y_{t}^{2}+4\lambda_{HS}+3\lambda_{S}\right)\\
\label{RGE-lamS}
Q\dfrac{d\lambda_{S}}{dQ} &=& \dfrac{1}{16\pi^{2}}(9\lambda_{S}^{2}+4\lambda_{HS}^{2})\\
\label{RGE-mH}
Q\dfrac{dm_H^{2}}{dQ} &=& \left(Q\dfrac{dm_H^{2}}{dQ} \right)_{SM} + \dfrac{1}{16\pi^{2}} \lambda_{HS}m_{S}^{2}\\
\label{RGE-mS}
Q\dfrac{d m_{S}^{2}}{dQ} &=& \dfrac{1}{16\pi^{2}}(3\lambda_{S}m_{S}^{2}+4\lambda_{HS}m_H^{2})
\end{eqnarray}
in which the RGEs of the gauge couplings $g_3$, $g_2$, $g_1$ and of the top Yukawa $y_t$ are implied. From $Q=m_S(m_S(M_G))$ down to $Q=m_h$, however, the working model is the SM and its parameters evolve as 
\begin{eqnarray}
\label{SM-RGE-lamH}
Q\dfrac{d\lambda_{H}}{dQ} &=& \dfrac{1}{16\pi^{2}}\left[\lambda_{H}(12\lambda_{H}-9g_{2}^{2}-3g_{1}^{2}+12y_{t}^{2}) + \dfrac{9}{4}g_{2}^{4}+\dfrac{3}{2}g_{1}^{2}g_{2}^{2}+\dfrac{3}{4}g_{1}^{4}-12y_{t}^{4}\right]\\
\label{SM-RGE-mH}
Q\dfrac{dm_H^{2}}{dQ} &=& \dfrac{1}{16\pi^{2}} m_H^{2}\left(6\lambda_{H}-\dfrac{9}{2}g_{2}^{2}-\dfrac{3}{2}g_{1}^{2} + 6y_{t}^{2} \right)
\end{eqnarray}
to reproduce the LHC results \cite{higgs} on the Higgs boson mass $m_h$ and the Higgs quartic coupling $\lambda_H$ (depending on the  input parameters at $Q=M_G$ and contribution of $S$ till $Q=m_S$). The illustration in Figure \ref{fig:RGEevol1} gives details of the two-stage RGE evolution. 

\begin{figure}[ht!]
\centering
\subfigure{\includegraphics[scale=1.15]{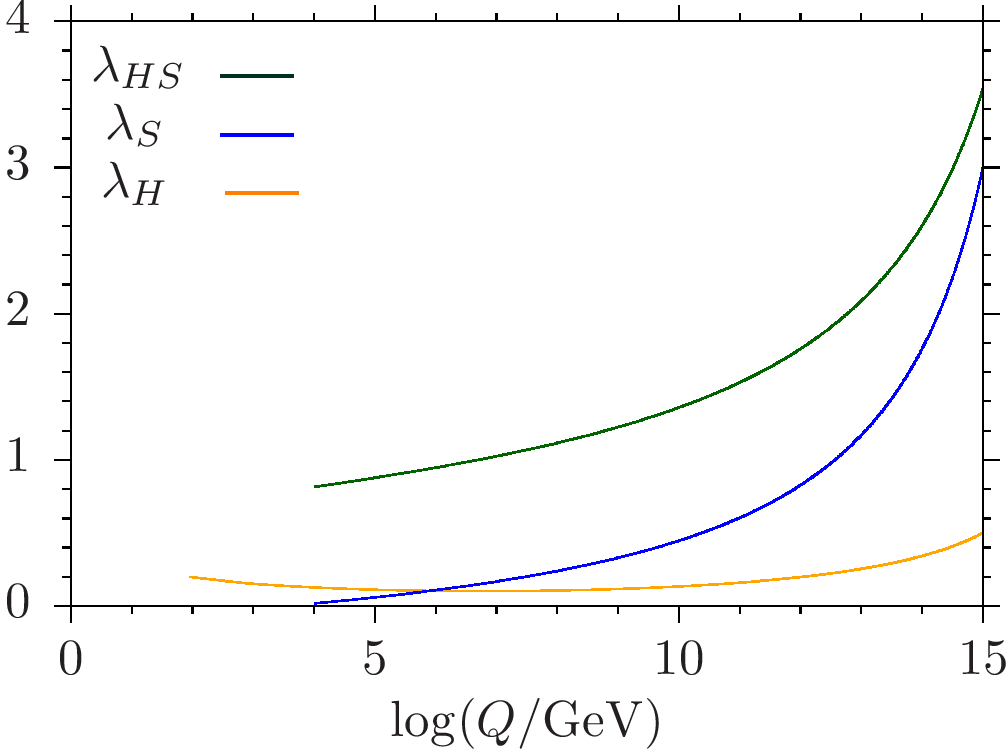}}%
\subfigure{\includegraphics[scale=1.15]{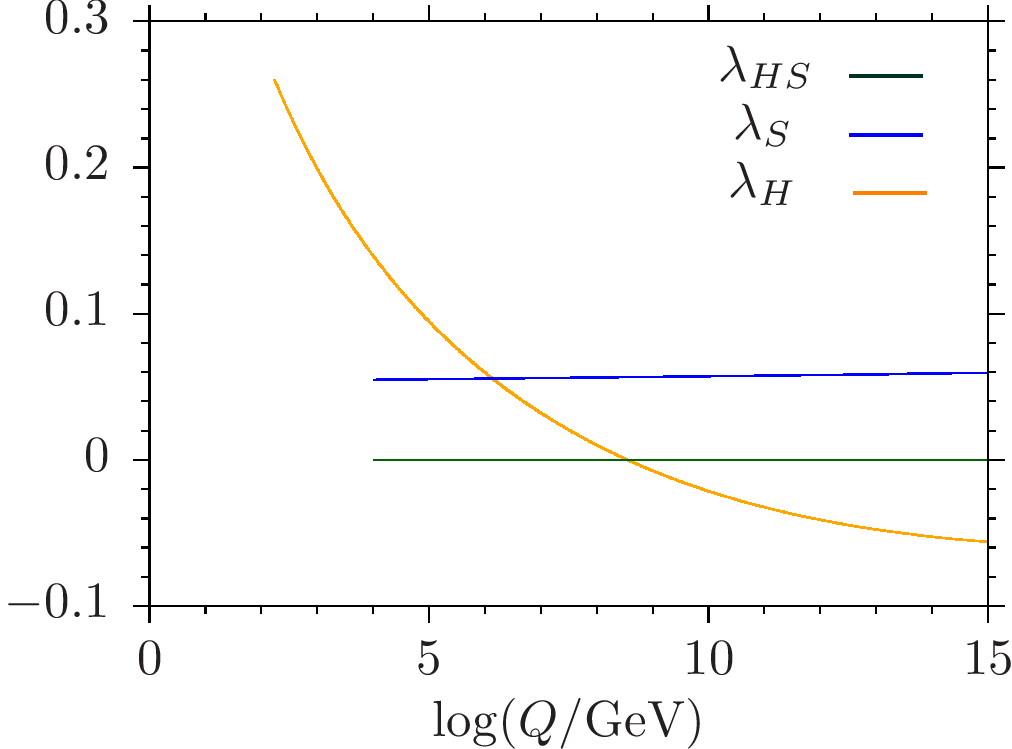}}
\caption{The RGE evolutions of $\lambda_H$, $\lambda_S$ and $\lambda_{HS}$ for $m_S(M_G)=10\ {\rm TeV}$. In the left panel, $\lambda_H(M_G)$, $\lambda_S(M_G)$ and $\lambda_{HS}(M_G)$ are assigned to avoid metastability. In the right panel, however, they are assigned in accordance with the MDDU linkup in (\ref{linkup}).}
\label{fig:lambda-lar}
\end{figure}

In what follows, we solve the RGEs to determine how a UV ($Q_0=M_G$) MDDU linkup like (\ref{linkup}) affects the Higgs boson mass and other observables. We start by analyzing the RGE evolutions of $\lambda_H(Q)$, $\lambda_S(Q)$ and $\lambda_{HS}(Q)$ for $m_S(M_G)=10\ {\rm TeV}$. The results are plotted in Fig. \ref{fig:lambda-lar}. In the left panel, as a beneficial approach, we consider those $\lambda_S(M_G)$ and $\lambda_{HS}(M_G)$ values (stretching them up to naive perturbative upper bound of $\sqrt{4\pi}$) for which the metastability \cite{meta} in the SM Higgs potential around $10^{11}\ {\rm GeV}$ is avoided. It is clear that the particular $\lambda_S(M_G)$ and $\lambda_{HS}(M_G)$ do indeed prevent the metastability ($\lambda_H(Q)$ remains positive for  the entire range).

In the right panel of Fig. \ref{fig:lambda-lar}, we set $\lambda_H(M_G)$, $\lambda_S(M_G)$ and $\lambda_{HS}(M_G)$ in accordance with the MDDU linkup (\ref{linkup}) (which takes the seesawic structure in (\ref{seesawic})). As suggested by its RGE in (\ref{RGE-lamHS}), $\lambda_{HS}(Q)$ remains small if it is small, that is, it is not generated by the loops if it is zero at the input scale. It is for this reason that  $\lambda_S(Q)$ and $\lambda_{SH}(Q)$ remain almost unchanged as they talk to the SM sector via $\lambda_{SH}$(Q). It is again for this reason that $\lambda_H(Q)$ (driven mainly by the top quark Yukawa $y_t(Q)$) nearly follows its SM evolution (with a slight slope change near $Q=m_S$). Thus, the MDDU linkup at the UV seems to be quite effective.

\begin{figure}[ht!]
\centering
\subfigure{\includegraphics[scale=1.14]{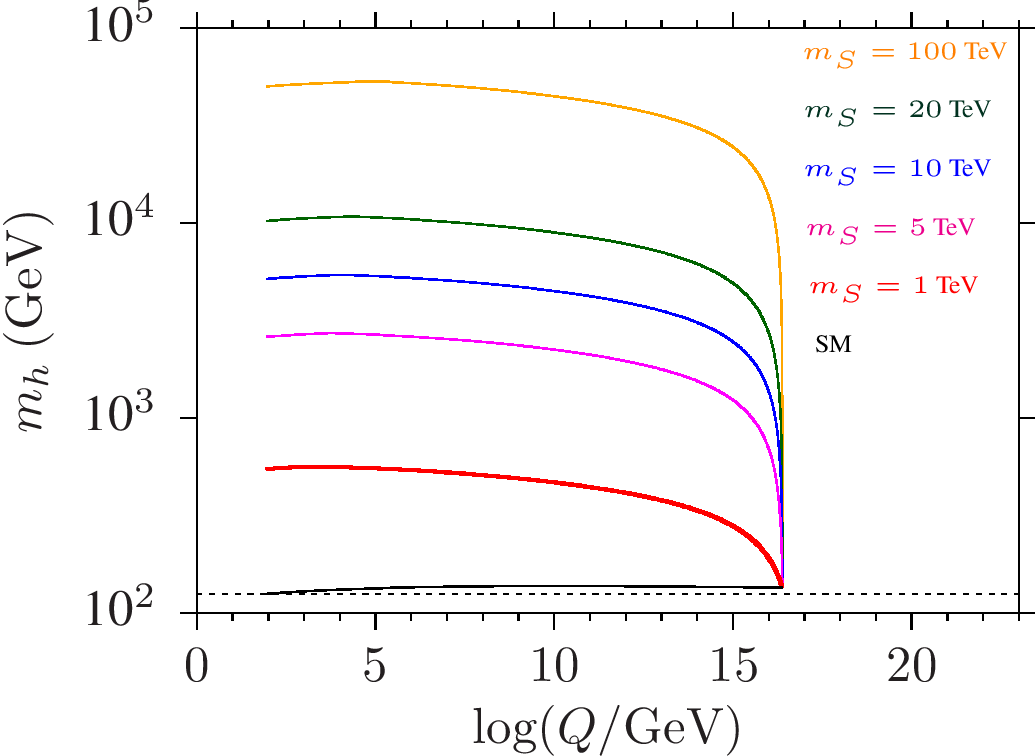}}%
\subfigure{\includegraphics[scale=1.15]{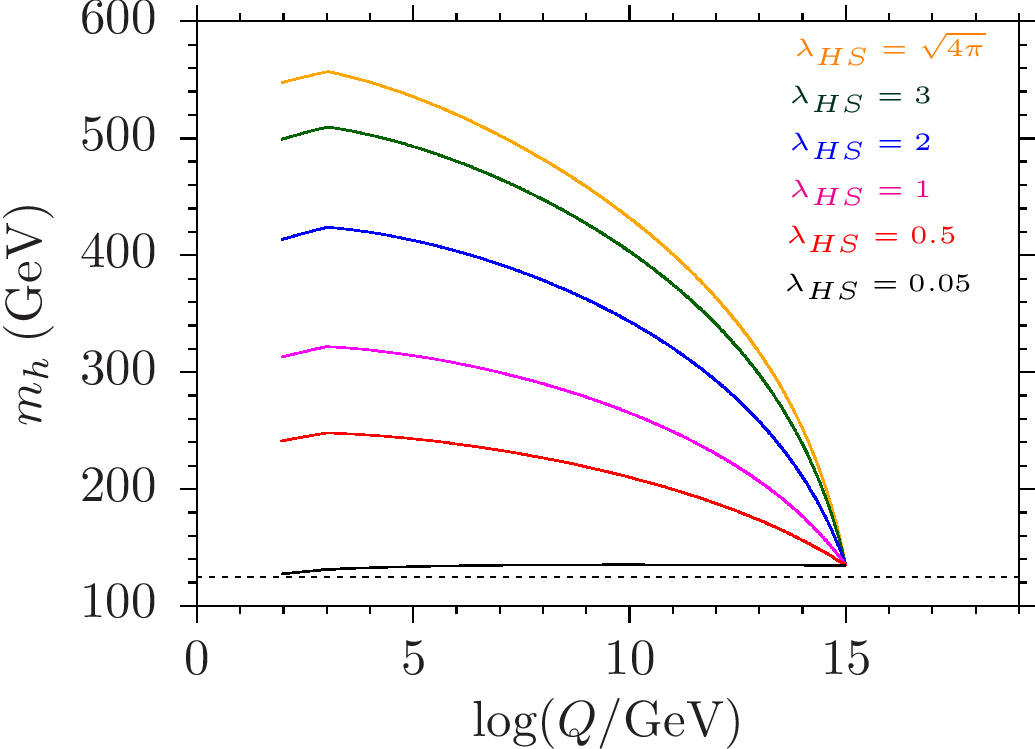}}
\caption{The RGE evolution of the Higgs boson mass $m_h(Q)$
for different $m_S(M_G)$ (left panel) and $\lambda_{HS}$ (right panel) values. (Fixed parameters are as in the left panel of Figure \ref{fig:lambda-lar}.)}
\label{fig:mh-ler}
\end{figure}

Having done with $\lambda_H(Q)$, $\lambda_S(Q)$ and $\lambda_{HS}(Q)$, we now turn to analysis of the Higgs boson mass $m_h(Q) = \sqrt{2} m_H(Q)$. The results are shown in Figure \ref{fig:mh-ler} in which the Higgs mass is plotted 
for different $m_S(M_G)$ (left panel) and $\lambda_{HS}$ (right panel) values. The parameters held fixed are as in the left panel of Figure \ref{fig:lambda-lar}. The left panel makes it clear that $m_h(m_h)$ overshoots the LHC result  
by orders of magnitude depending on how large $m_S(M_G)$ is. The reason is the quantum corrections in equation (\ref{corr-1}) or equivalently the second term at the right-hand side of its RGE in (\ref{RGE-mH}). This is the hierarchy problem. The Higgs boson mass grows to larger and larger values when it couples to a heavier and heavier scalar. 

The right panel of Figure \ref{fig:mh-ler}, on the other hand, makes it clear that smaller the $|\lambda_{HS}(M_G)|$ smaller the deviation of $m_h(m_h)$ from its LHC value. This actually is a clear proof that hierarchy problem can be alleviated only in UV completions which allow $\lambda_{HS}(M_G)$ 
to deviate from the SM couplings. The symmergence \cite{demir1,demir2} is one such completion in which $\lambda_{HS}(M_G)$ is allowed to even vanish, let alone the small values compared to $\lambda_H(M_G)$. 

\begin{figure}[ht!]
\centering
\includegraphics[scale=2.2]{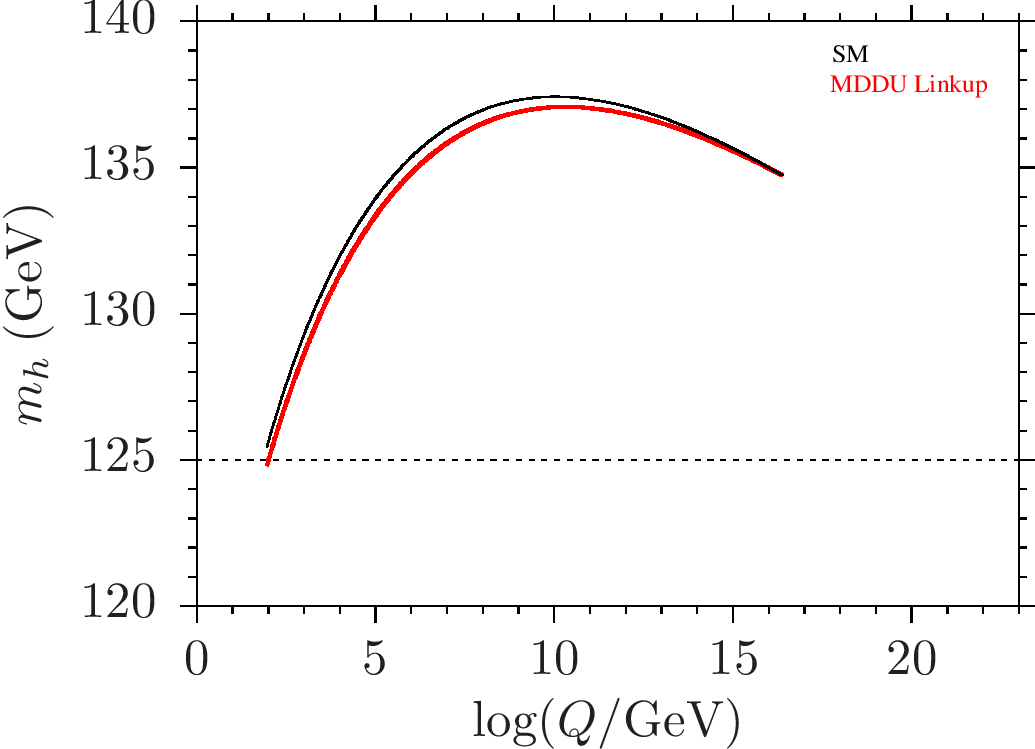}
\caption{The RGE evolution of the Higgs boson mass $m_h(Q)$ without (black curve) and with (red curve) the MDDU linkup (\ref{linkup}) at $Q_0=M_G$. The red curve contains individual curves for $m_S(M_G)=1$,5, 10, 20, 100 TeV, which nearly coincide to appear all red. This coincidence ensures that $m_h(Q)$ is indeed loosened from the $m_S(Q)$ thanks to the MDDU linkup (with the seesawic limit in (\ref{seesawic})).}
\label{fig:mh-linkup}
\end{figure}

Finally, we give in Figure \ref{fig:mh-linkup} the RGE evolution of the Higgs boson mass for the MDDU linkup (\ref{linkup}) at $Q_0=M_G$. It is clear that $m_h(Q)$ closely follows the SM evolution, and falls at $Q=m_h$ into the ballpark of the LHC result. This is the confirmation of the fact that a MDDU linkup of the form (\ref{linkup}) at $Q_0=M_G$ does indeed alleviate the hierarchy problem. The Higgs boson mass is made immune to the effects of the heavy scalars not at a specific scale like $Q=Q_0$ but at all the  scales beneath via the renormalization group flow. This happens thanks to the fact that $\lambda_{HS}$ remains small if it is small, as implied by its RGE in (\ref{RGE-lamHS}).

Having done with the Higgs boson mass, we turn our attention to vacuum energy. It shifts by the Coleman-Weinberg contribution 
\begin{equation}
\delta V = \dfrac{1}{64\pi^2} {\rm str}\left[{\mathcal{M}}^4 \log \left(\dfrac{{\mathcal{M}}^2}{e^{3/2}Q^2}\right)\right]
\label{eq:1Lscalarpotential}
\end{equation}
where the mass matrix ${\mathcal{M}}$ involves all the SM particles plus the singlet scalar $S$. It is expected to be ${\mathcal{O}}\left(m_S^4\right)$, which is gigantically large
compared to its observational value of order $m_\nu^4$ \cite{Planck}. We thus conclude that a MDDU linkup like (\ref{linkup}) saves the Higgs boson mass from the destabilizing UV effects but can simply do nothing about the cosmological constant problem. 

\section{Scalar Dark Matter with and without MDDU Linkup}
\label{sec:model}

In this section we analyze the parameter space of the model with the assumption that the singlet Scalar $S$ is considered to be a cold DM candidate. Our analysis examines two cases contrastively: A blind scan with no constraints except the perturbativity and the experimental bounds on the parameters, and a scan with the MDDU linkup (\ref{linkup}) at $Q_0=M_G$. It will be seen that the linkup reduces the parameter space considerably and reveals the regions where a scalar DM can be added to the SM in a natural way. 

In our scan, we accept only the solutions which yield a Higgs boson mass consistent with LHC's 125 GeV measurement up to some theoretical uncertainties. These uncertainties are dominated by those in the top quark mass and strong gauge coupling, which result in about 2 GeV uncertainty in theoretical calculations of the Higgs boson mass \cite{Degrassi:2002fi}. We set the top quark mass to its central value of 173.3 GeV \cite{Group:2009ad}, and keep in mind that $(1-2)\sigma$  variations in the top quark mass can shift the Higgs boson mass by $(1-2)$ GeV \cite{Gogoladze:2011aa}. Thus, we allow the Higgs boson mass in the range $(123-127)$ GeV after taking into account the loop contribution from the singlet scalar.

%In our scan, we accept only those solutions which yield a Higgs boson mass consistent with LHC's 125 GeV measurement, after including $\delta m_{h}^{2}$. In doing so, we allow for about 2 GeV uncertainty in calculation of the Higgs boson mass, dominated by the uncertainty in the top quark mass and strong gauge coupling. We set the top quark mass to its central value of 173.3 GeV \cite{Group:2009ad}, and keep in mind that $(1-2)\sigma$  variations in the top quark mass can shift the Higgs boson mass by $(1-2)$ GeV \cite{Gogoladze:2011aa}. To reiterate, we accept only the solutions that yield the Higgs boson mass in the range $(123-127)$ GeV after taking into account the loop contribution from the singlet scalar.

In our scans, we use the spectrum calculator SPheno-4.0.4 \cite{Porod:2003um} generated by SARAH-14.4.0 \cite{Staub:2013tta}. We further implement our scalar DM model into MicrOmegas-5.0.8 \cite{Belanger:2001fz} to calculate the DM observables. We constrain the parameter space with recent bounds from Planck measurements  within $5\sigma$ uncertainty \cite{Akrami:2018vks} as 
\begin{equation}
0.114 \leq \Omega h^{2} \leq 0.126
\label{eq:Planck}
\end{equation}
where $h$ is Hubble constant in units of today's value.  Note that $5\sigma$ uncertainty is considered to include the theoretical uncertainties in calculation of the DM relic density. Solving the Boltzmann equation evolving with the DM annihilation processes yields exponential dependence on the model parameters and the Higgs boson mass. A slight uncertainty in these parameters is, then, exponentially enhances in the relic density calculation (for details, see \cite{Belanger:2001fz}). 

We can summarize the parameters and their ranges as follows:
\begin{enumerate}
    \item {\it Blind Scan} in which we vary model parameters in the domain
\begin{eqnarray}
&&0 \leq  \lambda_{H}(Q_0)  \leq 1~, \nonumber\\
&&0 \leq  \lambda_{S}(Q_0)  \leq 1~, \nonumber\\
&&0 \leq  \lambda_{HS}(Q_0)  \leq 0.65~, \\
&&0 \leq  m_{S}(Q_0)  \leq 1~{\rm TeV},\nonumber
\label{eq:pspace}
\end{eqnarray}
accept only those points which agree with the experimental bounds such that $Q_0=M_G$ for 
$m_S \geq m_h$ and $Q_0=m_h$ for $m_S \leq m_h$.
 
\item {\it Linkup Scan} in which we vary model parameters in the domain
\begin{eqnarray}
&&0 \leq  \lambda_{H}(Q_0)  \leq 1~, \nonumber\\
&&0 \leq  \lambda_{S}(Q_0)  \leq 1~, \\
&&0 \leq  m_{S}(Q_0)  \leq 1~{\rm TeV},\nonumber
\label{eq:pspace}
\end{eqnarray}
accept only those points which agree with the experimental bounds such that $Q_0=M_G$ for 
$m_S \geq m_h$ and $Q_0=m_h$ for $m_S \leq m_h$
(after linking $\lambda_{HS}(M_G)$ to others as in the MDDU linkup (\ref{linkup})). 
\end{enumerate}
Note that we consider the perturbativity limit on $\lambda_{HS}$ in determining its range, and the experimental reach in $m_{S}$, while much heavier singlet scalar can still saturate the DM relic density given in (\ref{eq:Planck}). 

%Our analysis below will reveal implications of the Higgs mass stability is on the DM parameter domain (beside the general viability of the scalar DM model). 

\subsection{Higgs Profile}
\label{sec:Hprofile}

\begin{figure}[ht!]
\centering
\subfigure{\includegraphics[scale=1.2]{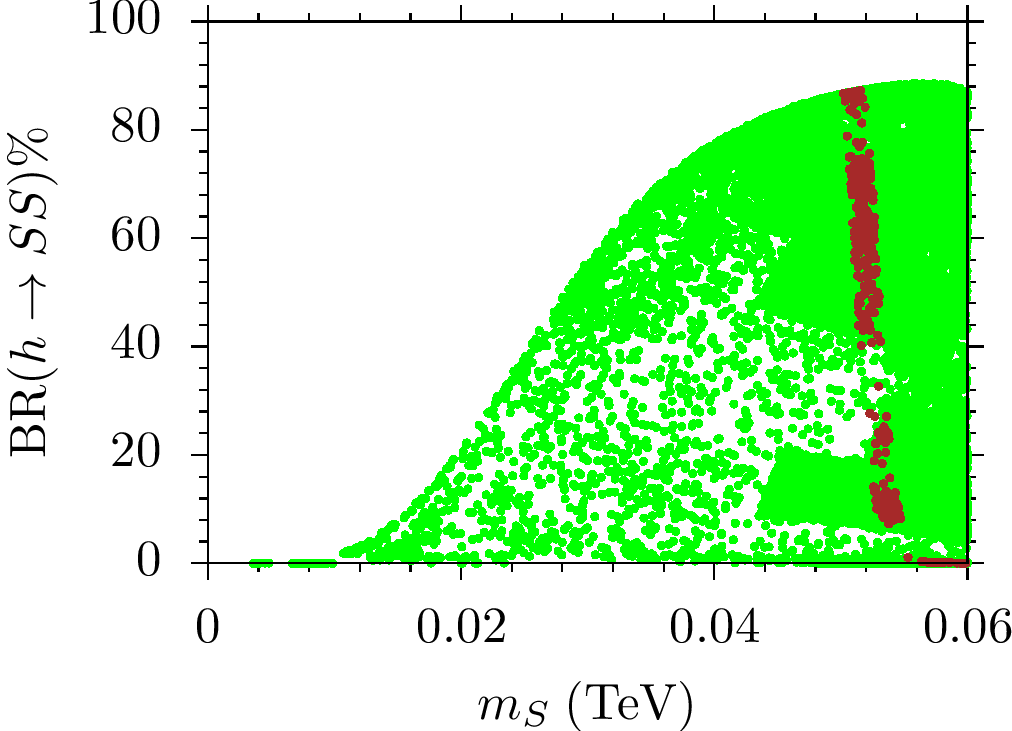}}%
\subfigure{\includegraphics[scale=1.2]{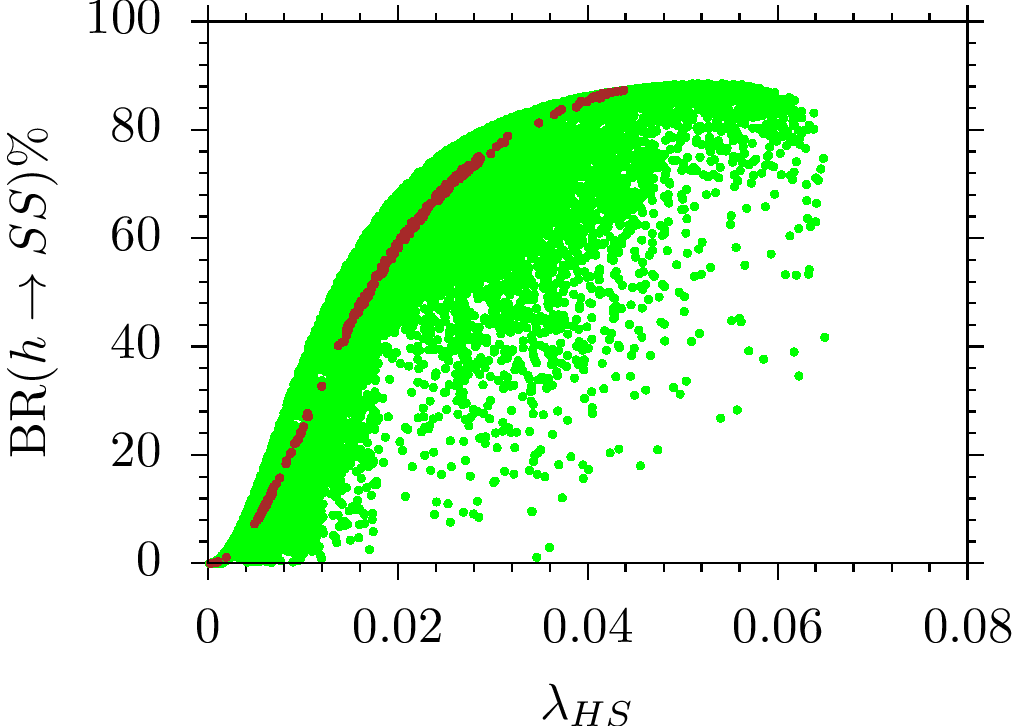}}
\caption{Plots for the Higgs boson decays in the ${\rm BR}{h\rightarrow SS}-m_{S}$, ${\rm BR}(h\rightarrow SS)-\lambda_{HS}$ planes. All points satisfy the Higgs boson mass constraint, while brown points are a subset of green and they are consistent with the Planck Satellite bound on relic abundance of the singlet scalar within $5\sigma$ uncertainty.}
\label{fig:Hdecays}
\end{figure}

\begin{figure}[ht!]
\centering
\subfigure{\includegraphics[scale=1.2]{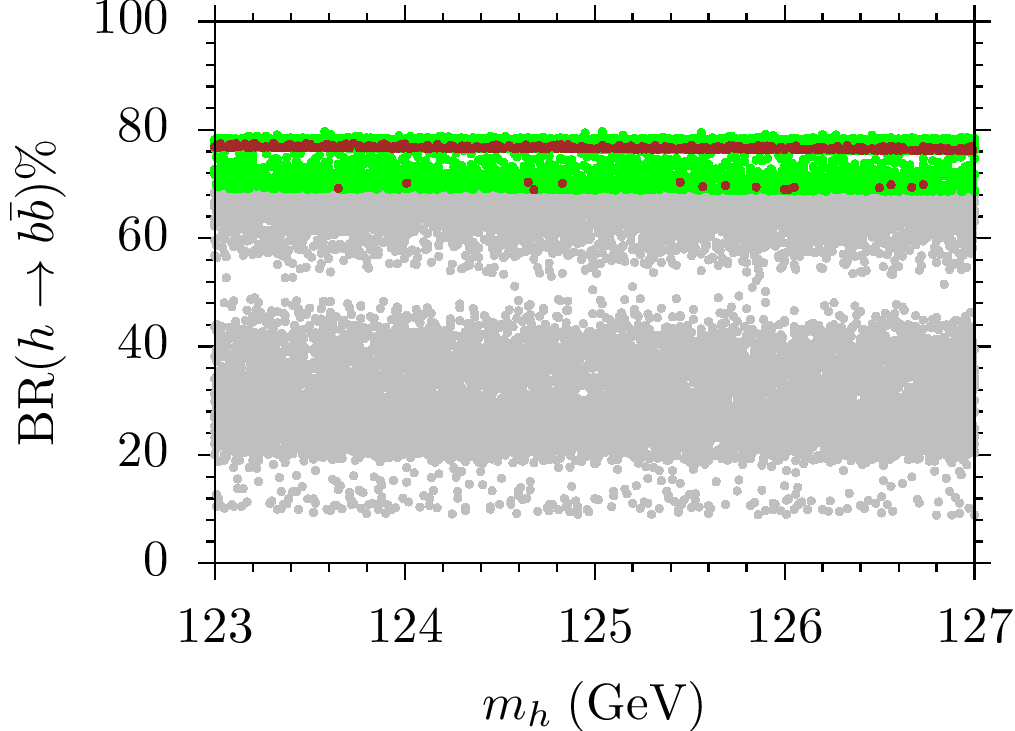}}%
\subfigure{\includegraphics[scale=1.2]{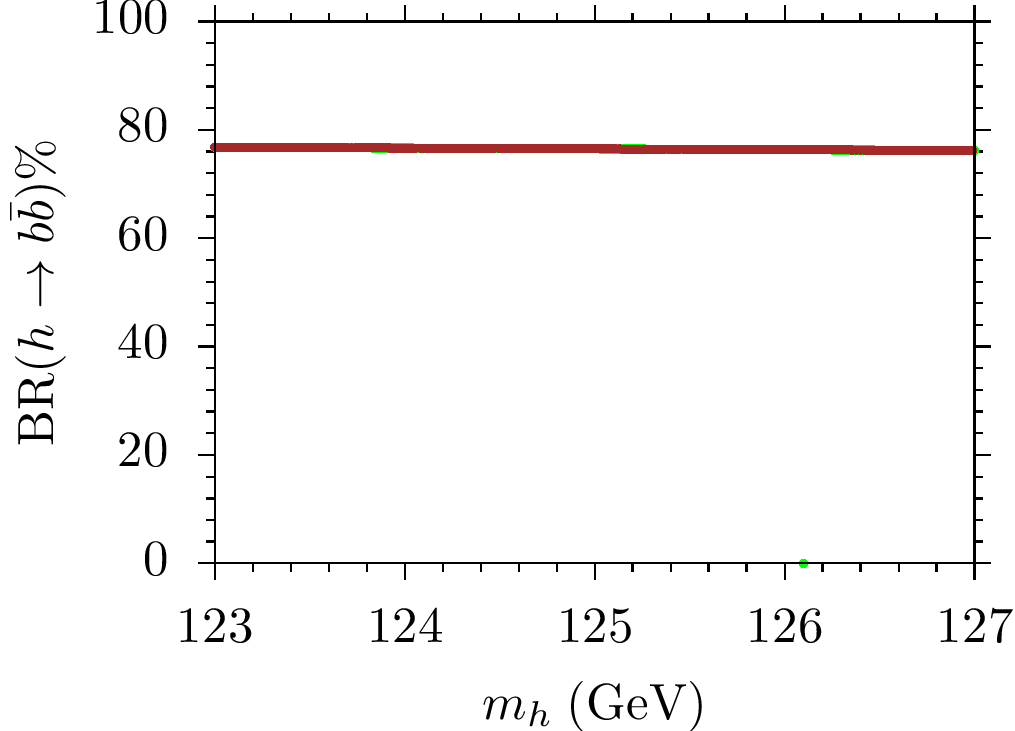}}
\subfigure{\includegraphics[scale=1.2]{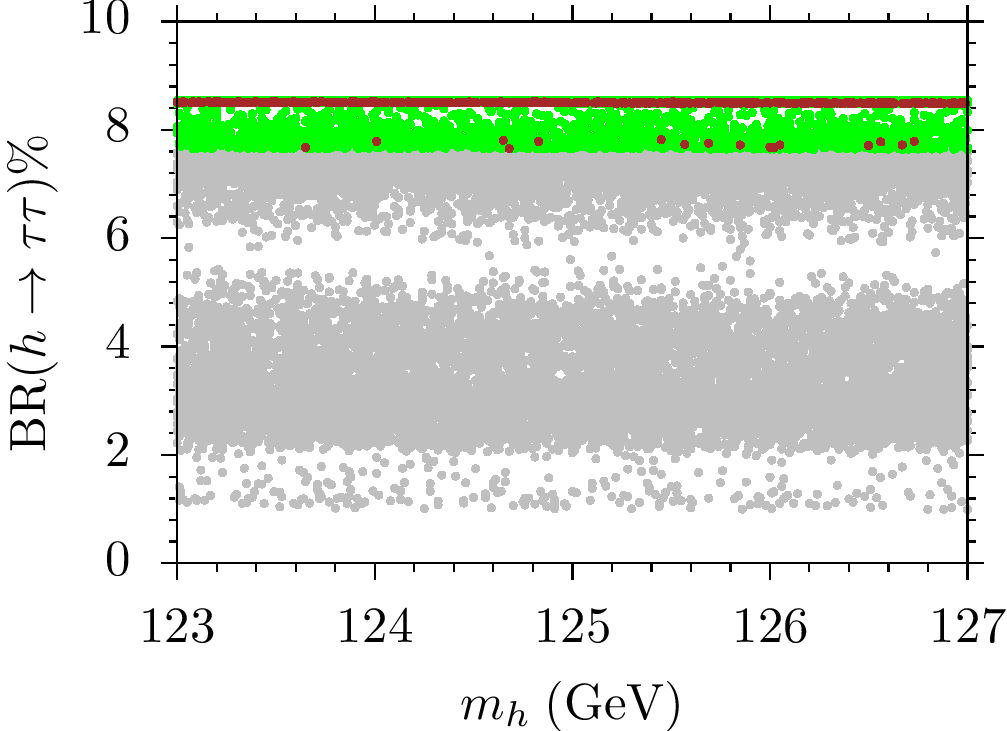}}%
\subfigure{\hspace{0.3cm}\includegraphics[scale=1.15]{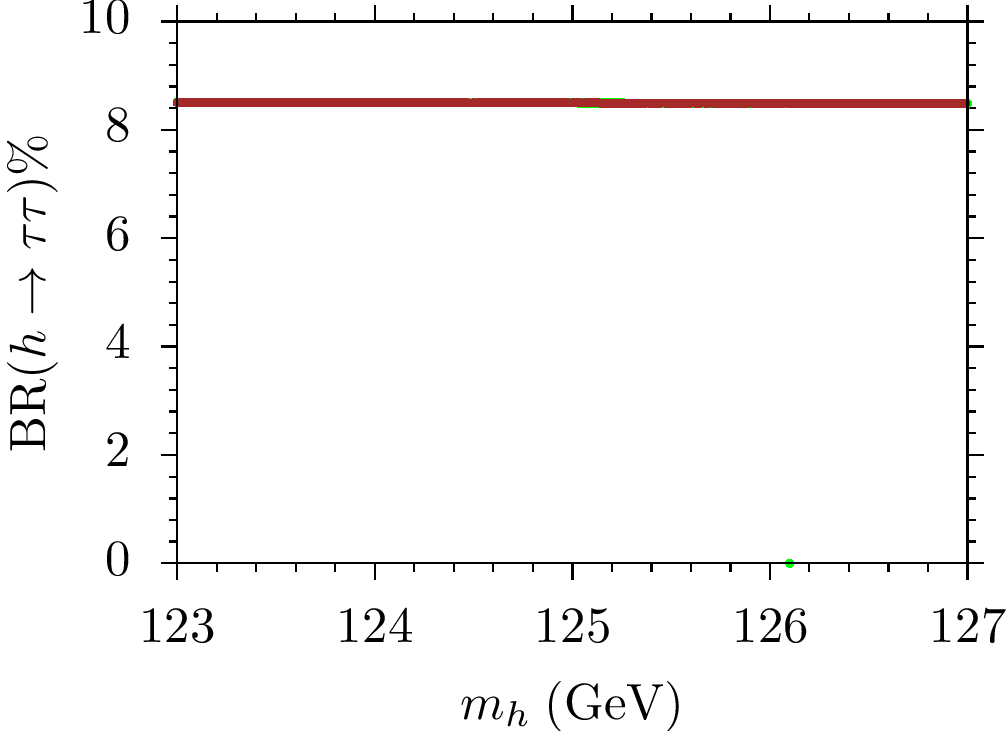}}
\caption{Plots for the Higgs boson decays in the ${\rm BR}(h\rightarrow bb)-m_{h}$ and ${\rm BR}(h\rightarrow \tau\tau)-m_{h}$ planes obtained from the blind scan (left) and MDDU linkup scan (right). All points satisfy the Higgs boson mass constraint. Gray points in the left panels are excluded by the large invisible Higgs boson decays, while the green points are allowed. Brown points form a subset of green, and they satisfy the Planck bound on the relic abundance of S.}
\label{fig:HdecaysSM}
\end{figure}

Even though we require the Higgs boson mass to be about 125 GeV, it can be assigned to be the SM-like Higgs boson if its decay modes are consistent with the SM predictions. The extra scalar field bring invisible decay modes of the Higgs boson when $m_{S} \lesssim m_{h}/2$. The constraint on the invisible decays of the Higgs boson vary depending on its production channels. Assuming the SM $ttH$ production cross-section the bound on the invisible decays can be set ${\rm BR}(h\rightarrow {\rm invisible}) < 0.46$ \cite{CMS:2019bke}, while it is ${\rm BR}(h\rightarrow {\rm invisible}) < 0.37$ if the Higgs boson is produced through the vector boson fusion processes \cite{Aaboud:2018sfi}. Furthermore, a combination of direct searches yield a bound as ${\rm BR}(h\rightarrow {\rm invisible}) \lesssim 0.25$ \cite{Sirunyan:2018owy}. The invisible decay rates of the Higgs boson below $25\%$ do not change the dark matter implications \cite{Djouadi:2011aa}, Recovering the SM predictions we will require the solutions to satisfy ${\rm BR}(h\rightarrow SS)\leq 10\%$.

The decay width for the Higgs boson decaying into a pair of the singlet scalars is given for $2m_{S}\leq m_{h}$ as \cite{Belanger:2013xza}

\begin{equation}
\Gamma(h\rightarrow SS) = \dfrac{\lambda_{HS}^{2}v_{SM}^{2}}{64\pi^{2}m_{h}}\left( 1-\dfrac{4m_{S}^{2}}{m_{h}^{2}}\right)^{1/2}
\label{eq:invwidth}
\end{equation}

Figure \ref{fig:Hdecays} shows the plots for the invisible Higgs boson decays obtained through the blind scan in correlation with $m_{S}$ and $\lambda_{HS}$. All points satisfy the Higgs boson mass constraint, while green points Brown points are a subset of green and they are consistent with the Planck Satellite bound on relic abundance of the singlet scalar within $5\sigma$ uncertainty. the Higgs boson can decay into a pair of singlet scalars up to about $80\%$, when the scalar mass is realized near the resonance region ($m_{S} \simeq m_{h}/2$), when $\lambda_{HS}\sim 0.045$. The ${\rm BR}(h\rightarrow SS)-\lambda_{HS}$ plane shows that one can satisfy the condition on the invisible decay rates if $\lambda_{HS}\lesssim 0.08$ in this region without excluding any solution with $m_{S}\lesssim m_{h}/2$. 

A call for small $\lambda_{HS}$ can be realized when one imposed the MDDU linkup at the GUT scale, whose results are showin in Figure \ref{fig:HdecaysSM} in comparison with those from the blind scan side by side. All points satisfy the Higgs boson mass constraint. Gray points in the left panels are excluded by the large invisible Higgs boson decays, while the green points are allowed. Brown points form a subset of green, and they satisfy the Planck bound on the relic abundance of S. A general scan predicts branching ratios for the Higgs boson decays in a range wider than the SM results. Even though excluding large invisible decays of the Higgs boson narrows down the range, there are still solutions below the SM predictions as shown in the left panels for the $h\rightarrow bb$ and $h\rightarrow \tau\tau$. On the other hand, as can be seen from the right panels, imposing linkup condition removes all the predictions except those of the SM. Linkup predicts ${\rm BR}(h\rightarrow bb) \sim 78\%$ and ${\rm BR}(h\rightarrow \tau\tau) \sim 9\%$, which overlap with the SM predictions.

%Figure \ref{fig:Hdecays} shows the plots for the Higgs boson decays in the ${\rm BR}{h\rightarrow SS}-m_{S}$, ${\rm BR}(h\rightarrow SS)-\lambda_{HS}$, ${\rm BR}(h\rightarrow bb)-m_{h}$ and ${\rm BR}(h\rightarrow \tau\tau)-m_{h}$ planes. All points satisfy the Higgs boson mass constraint, while green points Brown points are a subset of green and they are consistent with the Planck Satellite bound on relic abundance of the singlet scalar within $5\sigma$ uncertainty. In the bottom panels, green points are excluded by the constraint on the invisible decays of the Higgs boson. We can see from the top panels that the Higgs boson can decay into a pair of singlet scalars up to about $80\%$, when the scalar mass is realized near the resonance region ($m_{S} \simeq m_{h}/2$), when $\lambda_{HS}\sim 0.045$. The ${\rm BR}(h\rightarrow SS)-\lambda_{HS}$ plane shows that one can satisfy the condition on the invisible decay rates if $\lambda_{HS}\lesssim 0.08$ in this region without excluding any solution with $m_{S}\lesssim m_{h}/2$. The bottom panels of Figure \ref{fig:Hdecays} exemplify the Higgs boson decays into SM particles. If we exclude the solutions with ${\rm BR}(h\rightarrow SS) > 10\%$, then we realize ${\rm BR}(h\rightarrow bb) \sim 80\%$ and ${\rm BR}(h\rightarrow \tau \tau)\sim 9\% $, which more or less overlap with the SM predictions.

\subsection{Dark Matter Implications}
\label{sec:DM}

In this section we consider the dark matter implications of the model in details. We first discuss the results from the general analyses which takes $\lambda_{HS}$ to be a free parameter, while we employ the relation given in Eq.(\ref{linkup}) and emphasize its exclusiveness in regard of the dark matter phenomenology.

\begin{figure}[ht!]
\centering
\subfigure{\includegraphics[scale=1.2]{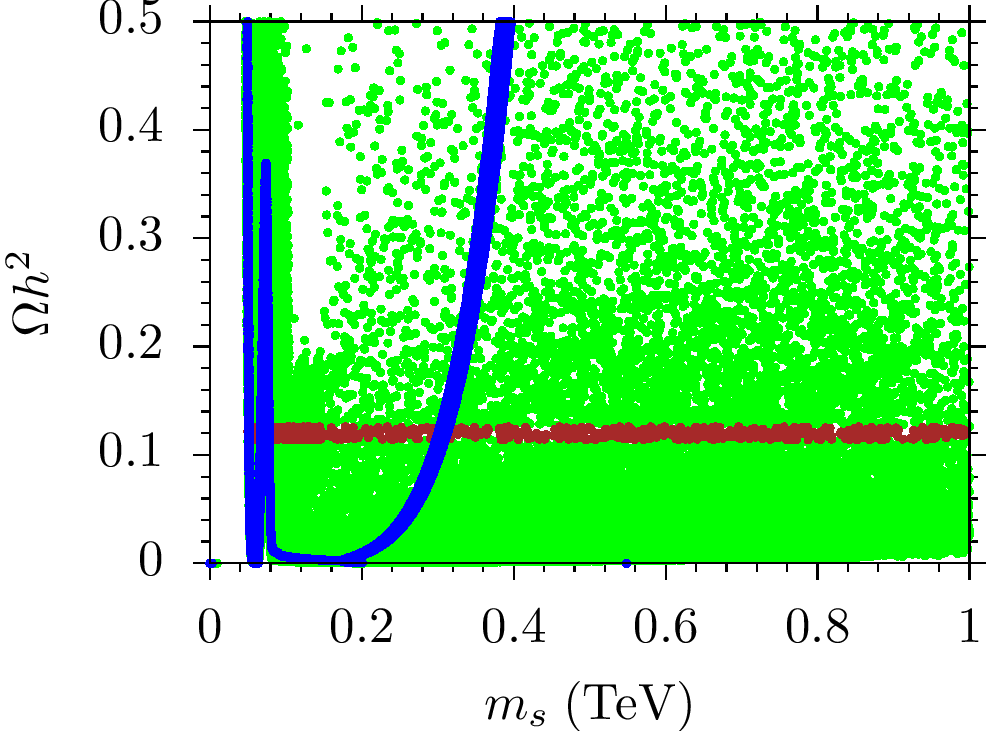}}%
\subfigure{\includegraphics[scale=1.2]{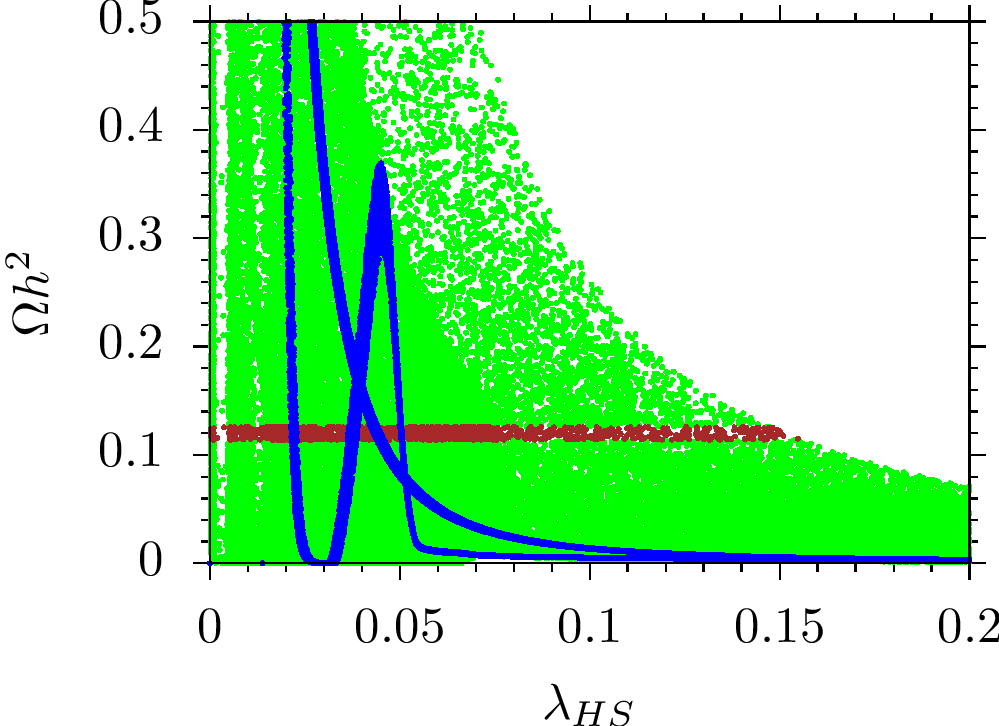}}
%\subfigure{\includegraphics[scale=1.2]{SDMfixL_oh2_mS.png}}%
%\subfigure{\includegraphics[scale=1.2]{SDMfixL_oh2_LamSH.png}}
\caption{The relic abundance of the singlet scalar in correlation with its mass (left) and its coupling (right) to the Higgs boson. All points are consistent with the Higgs boson mass constraint. Green points are also compatible with the condition from the invisible decay rates of the Higgs boson. Blue points form a subset of green, and they represent the solutions in which $\lambda_{HS}$ is fixed with MDDU linkup (\ref{linkup}). Brown points are another subset of green, and they are allowed by the current Planck bound on the relic abundance of $S$ within $5\sigma$ uncertainty. The blue points satisfy the Planck bound on the relic abundance when they cross the brown region.}
\label{fig:relic}
\end{figure}

Figure \ref{fig:relic} displays our results for the relic abundance of $S$ in correlation with its mass (left) and its coupling (right) to the Higgs boson. All points are consistent with the Higgs boson mass constraint. Green points are also compatible with the condition from the invisible decay rates of the Higgs boson. Blue points form a subset of green, and they represent the solutions in which $\lambda_{HS}$ is fixed with the MDDU linkup (\ref{linkup}). Brown points are another subset of green, and they are allowed by the current Planck bound on the relic abundance of $S$ within $5\sigma$ uncertainty. The blue points satisfy the Planck bound on the relic abundance when they cross the brown region. The brown points in the $\Omega h^{2}-m_{S}$ plane show that the correct relic abundance of $S$ can be realized at any mass of $S$, while it becomes quite exclusive when $\lambda_{HS}$ is fixed as given in the MDDU linkup (\ref{linkup}). The blue points reveal a unique correlation between $\Omega h^{2}$ and $m_{S}$, and the correct relic abundance can be satisfied when $m_{S}\simeq 50$, $80$ and $300$ GeV. The first two mass scales of $S$ correspond to the resonance region in which two dark matter particles annihilate into a Higgs boson.

A similar discussion can be followed for the right panel of Figure \ref{fig:relic}. The $\Omega h^{2}-\lambda_{HS}$ panel shows that the Planck bound on the relic abundance can be satisfied in the region with $\lambda_{HS} \lesssim 0.16$, while $S$' relic abundance is realized smaller than the lower bound from the Planck measurements for $\lambda_{HS}\gtrsim 0.16$. Indeed, the relic abundance over most of the fundamental parameter space is found lower than the Planck bound, and this region can be available if another sector for the dark matter is proposed. As seen from the blue points, the MDDU linkup (\ref{linkup}) yields two correlations between $\Omega h^{2}$ and $\lambda_{HS}$. The curve with extrema corresponds to the region with $m_{S}^{2} < |m_H^{2}|$, while the second curve behaves like an exponential correlation for $m_{S}^{2} \geq |m_H^{2}|$. 

\begin{figure}[t!]
\centering
\subfigure{\includegraphics[scale=1.2]{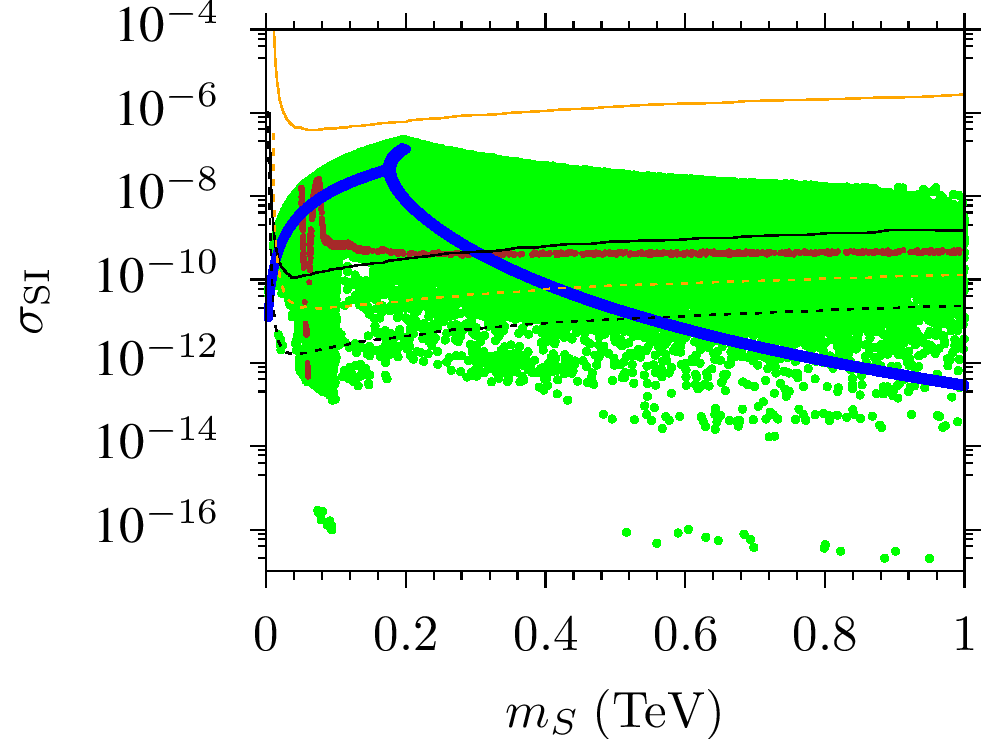}}%
\subfigure{\includegraphics[scale=1.2]{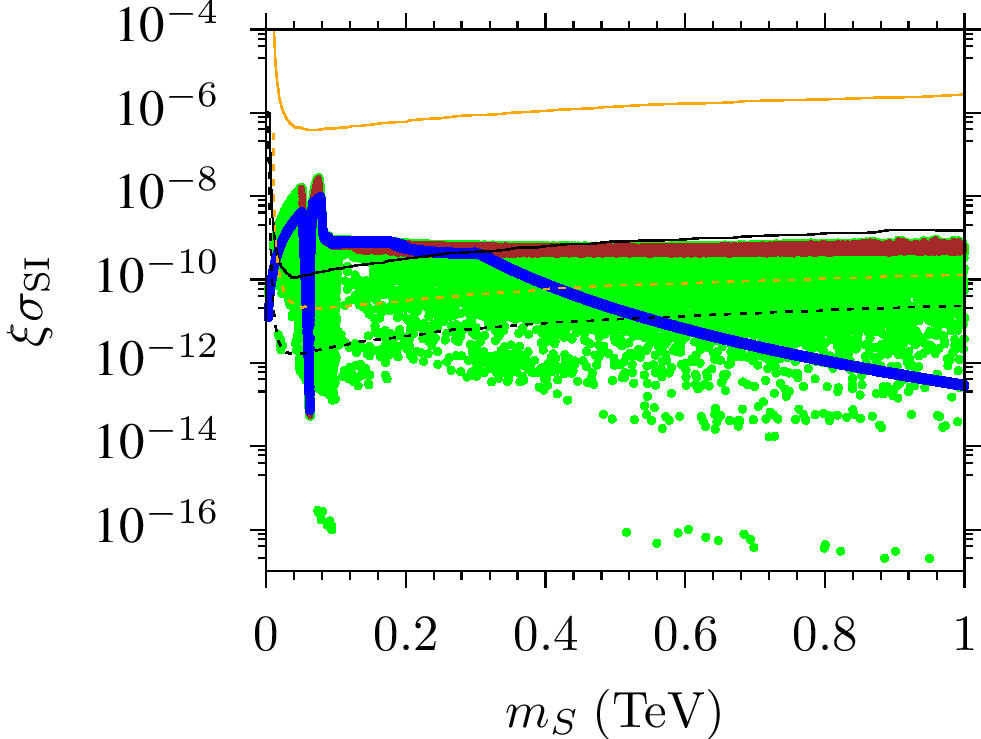}}
%\subfigure{\includegraphics[scale=1.2]{SDMfixL_SI_mS.png}}%
%\subfigure{\includegraphics[scale=1.2]{SDMfixL_SI_mS_2.png}}
\caption{The spin independent scattering cross-section of the dark matter. The color coding is the same as Figure \ref{fig:relic}. In addition, the brown points in the right panel represent the solutions with $0 \leq \Omega h^{2}\leq 0.126$. The orange solid (dashed) curve represents the current (future projection of) the exclusion from the Super-CDMS experiment \cite{Brink:2005ej}, and the black solid (dashed) curve displays the current (future) bounds from the LUX-Zeplin experiment \cite{Akerib:2018dfk}.}
\label{fig:SI}
\end{figure}

We continue the discussion of dark matter phenomenology with the spin-independent scattering of the DM at nuclei as shown in Figure \ref{fig:SI}. The color coding is the same as Figure \ref{fig:relic}. In addition, the brown points in the right panel represent the solutions with $0 \leq \Omega h^{2}\leq 0.126$. The orange solid (dashed) curve represents the current (future projection of) the exclusion from the Super-CDMS experiment \cite{Brink:2005ej}, and the black solid (dashed) curve displays the current (future) bounds from the LUX-Zeplin experiment \cite{Akerib:2018dfk}.

The left panel shows that the current experimental bounds on the spin-independent scattering of the dark matter have a strong impact on the parameter space of the model such that the scattering cross-section can be consistent with the experimental results when $m_{S}\gtrsim 250$ GeV. Besides, this region lies slightly below the current experimental bound from the LUX-Zeplin experiment, and one can expect this region to be tested very soon. In addition, it is also possible to realize testable allowed results in the resonance region ($m_{S} \sim m_{h}/2$). However, the solutions with large scattering cross-sections (in green) mostly correspond to those with relic abundance lower than the Planck bound, and as previously discussed, these solutions can be available when another sector for dark matter is proposed. When its relic abundance is lower than the Planck measurements (i.e. $\Omega h^{2} < 0.114$), it can constitute only a fraction of dark matter proportional to $\xi$ \cite{Belanger:2015vwa} as 

\begin{equation}
\setstretch{2.5}
\sigma \rightarrow \xi\sigma~,\hspace{0.3cm} {\rm where}\hspace{0.3cm} \xi = \left\lbrace \begin{array}{ll}
1 & {\rm for~} 0.114 \leq \Omega h^{2} \leq 0.126 \\
\dfrac{\Omega h^{2}}{0.12} & {\rm for}~ \Omega h^{2} < 0.114 
\end{array}\right.
\label{eq:xi}
\end{equation}
where $0.12$ is the central value of the Planck bound. If $S$ does not fully form the dark matter, it cannot fully account for the dark matter observations. The right panel shows also these solutions in brown, and as can be seen, almost all solutions with large scattering cross-sections disappear. In addition, the resonance region also reveals results, which are likely expected to be tested in near future. 

The blue points are obtained when we fixed $\lambda_{HS}$ as in the MDDU linkup (\ref{linkup}). A fixed $\lambda_{HS}$ yields a unique correlation also for the scattering cross-section as seen from the left panel. In this case, only the solutions with $m_{S}\sim 300$ GeV can be consistent both with the Planck bound and the LUX-Zeplin exclusion. In addition, these solutions are expected to be tested very soon. If we also involve the solutions with low relic abundance of $S$, then the Higgs boson resonance region also provides consistent and testable solutions in near future, as shown with the blue points in the right panel of Figure \ref{fig:SI}.

\begin{figure}[t!]
\centering
\subfigure{\includegraphics[scale=0.8]{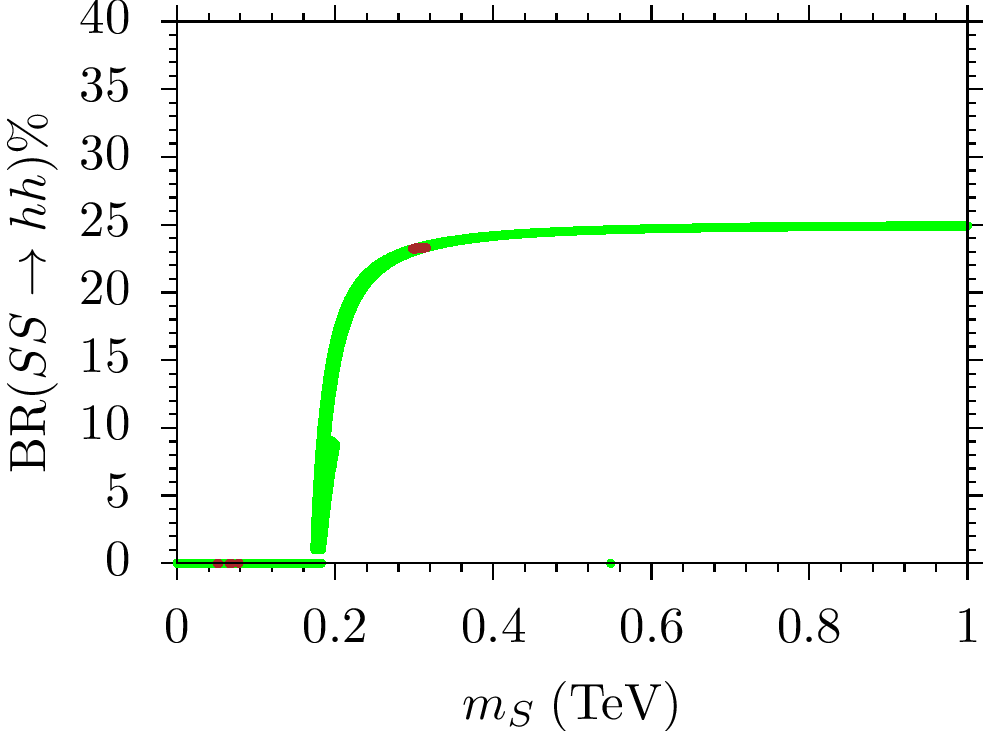}}%
\subfigure{\includegraphics[scale=0.8]{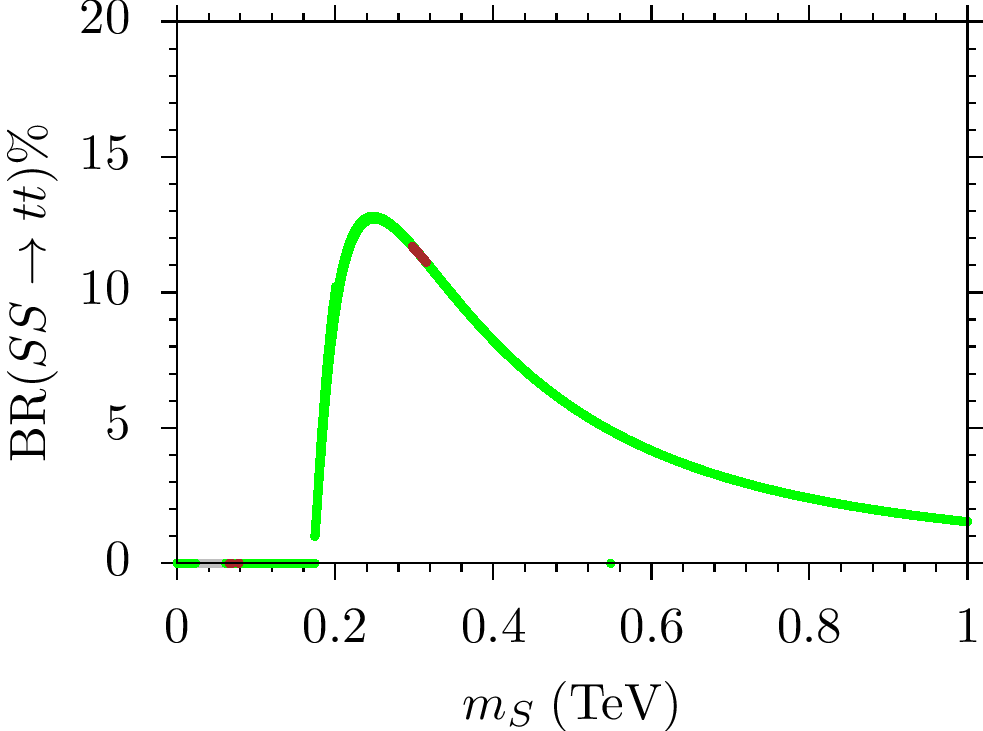}}%
\subfigure{\includegraphics[scale=0.8]{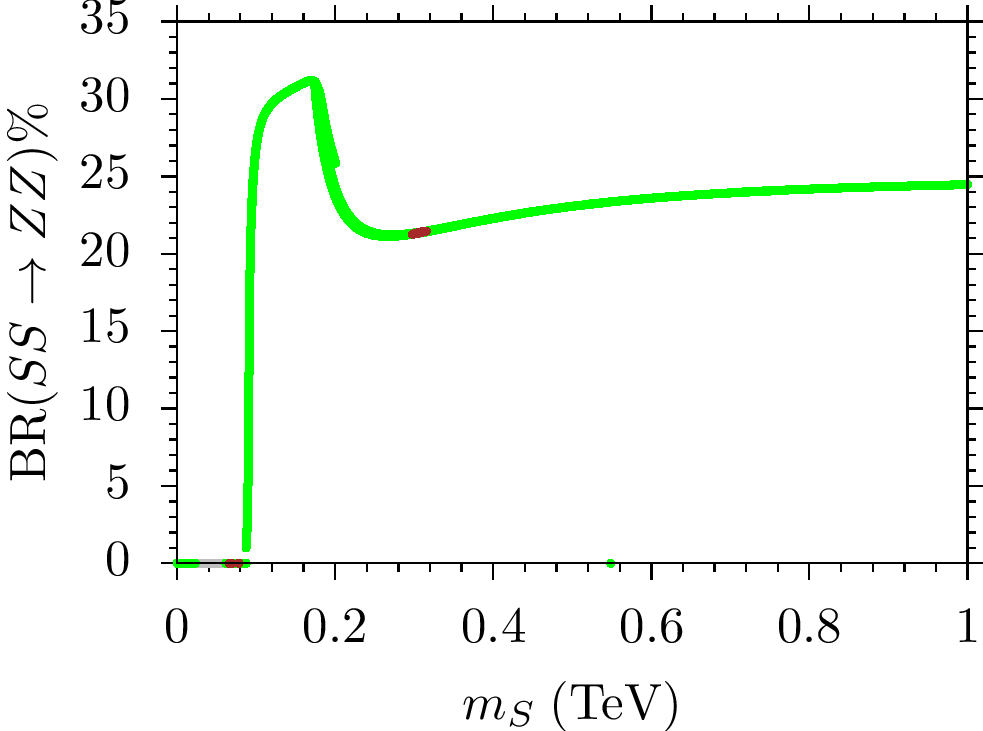}}
\subfigure{\includegraphics[scale=0.8]{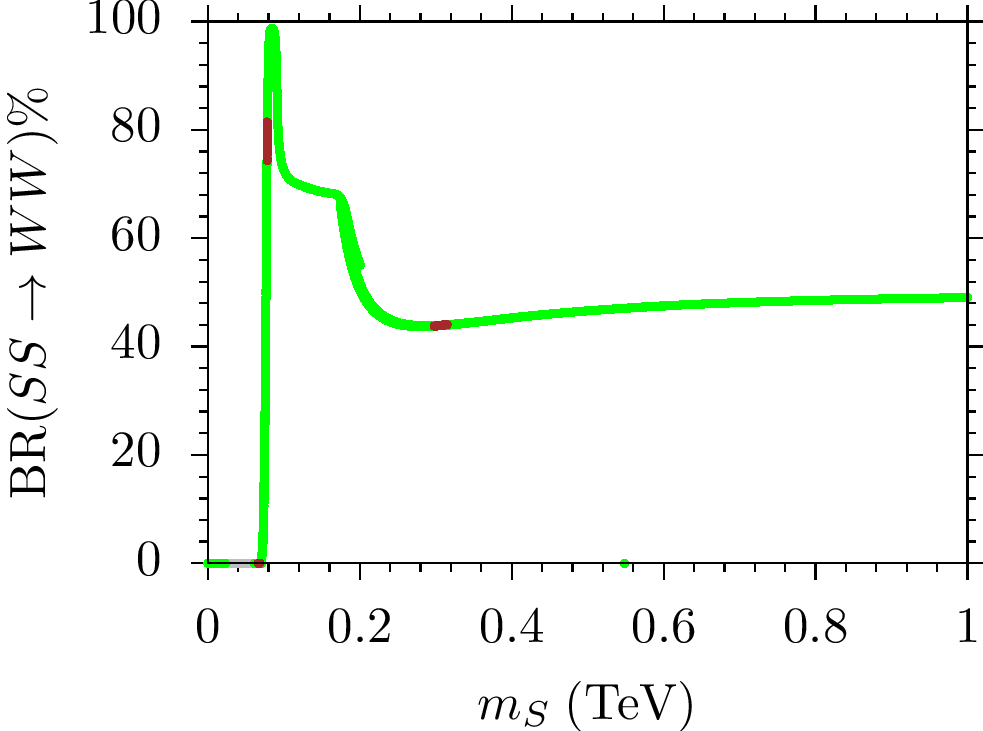}}%
\subfigure{\includegraphics[scale=0.8]{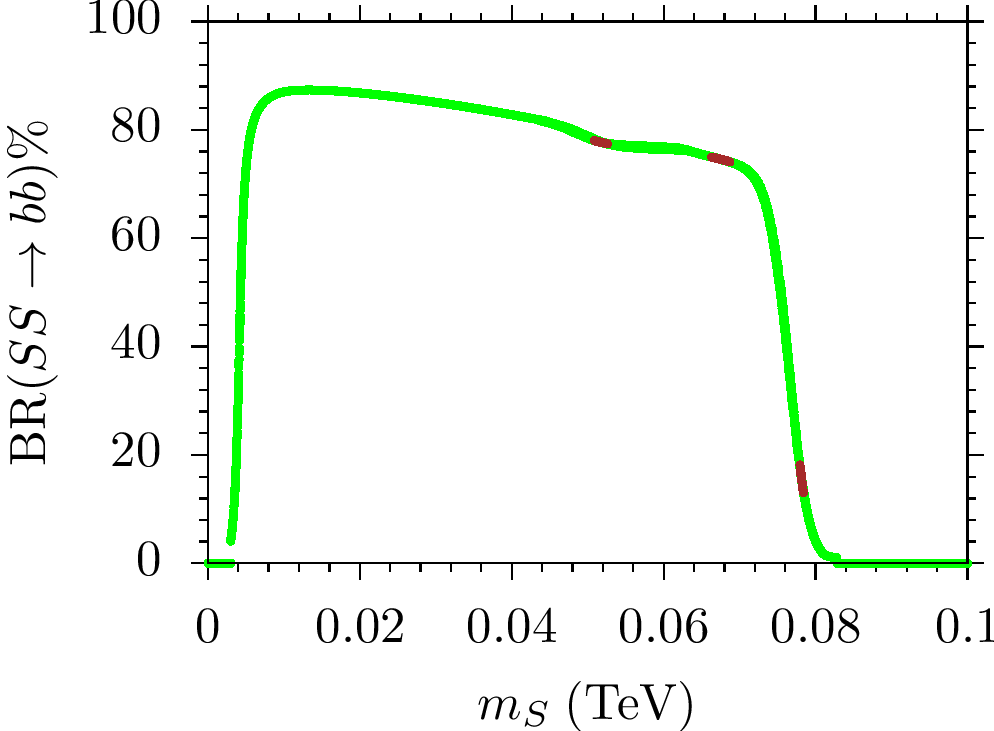}}%
\subfigure{\includegraphics[scale=0.8]{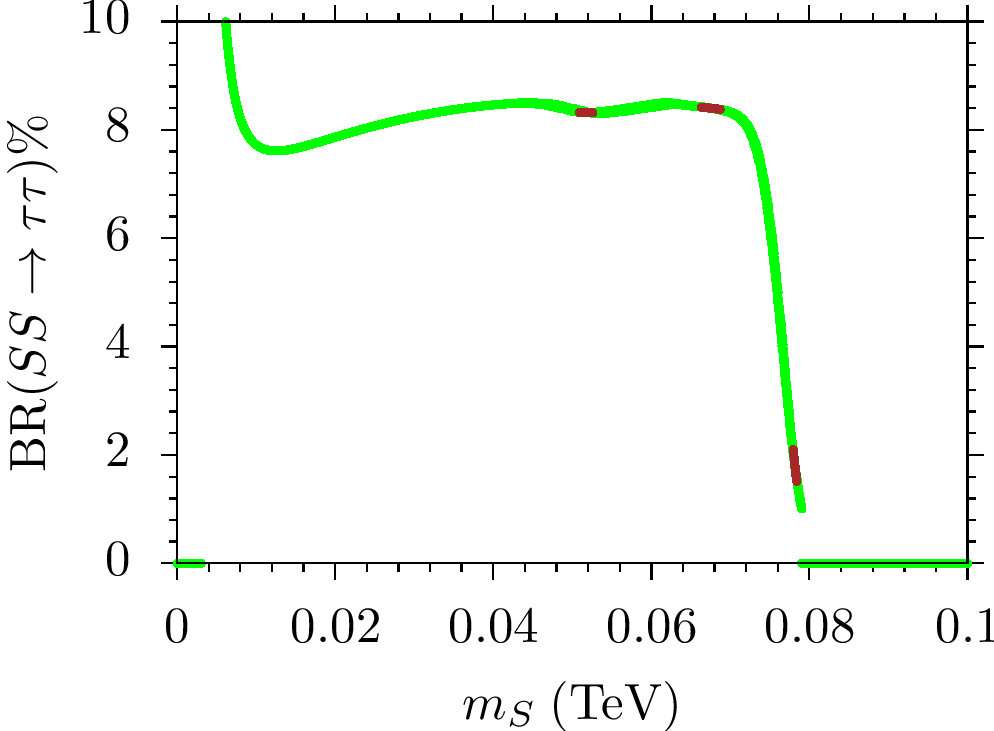}}

%\subfigure{\includegraphics[scale=1.2]{SDMBRSStt_mS.png}}%
%\subfigure{\includegraphics[scale=1.2]{SDMBRSStt_LamSH.png}}
\caption{The annihilation channels of the dark matter. The color coding is the same as Figure \ref{fig:relic}.}
\label{fig:annihilation}
\end{figure}

Since we consider only one field forming the dark matter, its annihilation processes take important part in realizing the correct relic abundance. Even though the linkup scan allows a very narrow range for the solutions consistent with the DM constraints, it predicts the same annihilation channels as those realized in the blind scan. Thus, we show the results only from the linkup scan in Figure \ref{fig:annihilation}. The rates of annihilation channels, in most cases, exhibit a peak when $m_{S}$ is close by the SM particle participating in the annihilation processes. According to the results, $SS\rightarrow hh$ can be realized at a rate about $25\%$. When $m_{S}\sim m_{t}$, $SS\rightarrow tt$ processes happen at about $13\%$, while its rate drops below $5\%$ for $m_{S}\gtrsim 500$ GeV. Another annihilation channel involves with a pair of $Z-$boson and it can be realized at about $30\%$ when $m_{S}\sim M_{Z}$. However, DM constraints (brown points) allow this processes to happen at about $22\%$.  An interesting possible annihilation channel of $S$ is one in which it annihilates into a pair of $W-$boson. The DM constraints allow this processes up to about $80\%$ when $m_{S}\sim M_{W}$, while it can also hapen at about $40\%$ when $m_{S}\sim 300$ GeV. One can identify also the channels involving a pair of $b-$quark and $\tau-$leptons, which take part when $m_{S} < m_{h}$. $SS\rightarrow bb$ and $SS\rightarrow \tau\tau$ processes can be realized at about $80\%$ and $8\%$, respectively.

%\begin{figure}[ht!]
%\centering
%\subfigure{\includegraphics[scale=1.2]{SDMBRSSbb_mS.png}}%
%\subfigure{\includegraphics[scale=1.2]{SDMBRSStautau_mS.png}}
%\subfigure{\includegraphics[scale=1.2]{SDMBRSSWW_mS.png}}%
%\subfigure{\includegraphics[scale=1.2]{SDMBRSStt_mS.png}}
%\caption{More annihilation channels of $S$ into a pair of bottom quarks, tau leptons, $W$ bosons and top quarks. The color coding is the same as Figure \ref{fig:relic}.}
%\label{fig:annihilation2}
%\end{figure}

\begin{figure}[t!]
\centering
\subfigure{\includegraphics[scale=1.2]{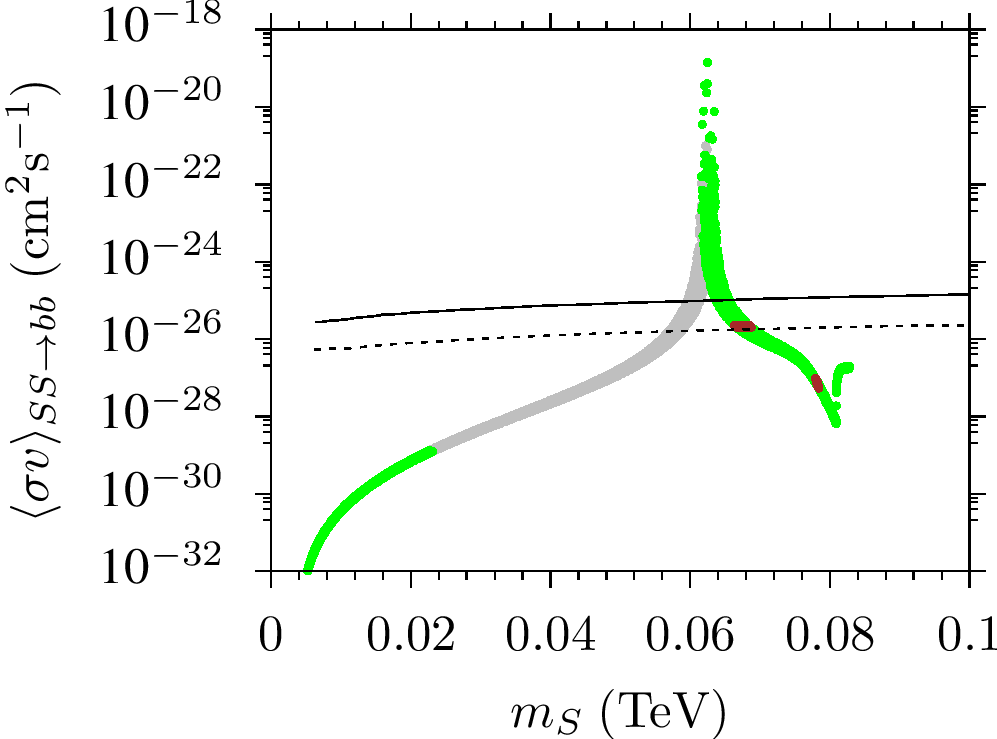}}%
\subfigure{\includegraphics[scale=1.2]{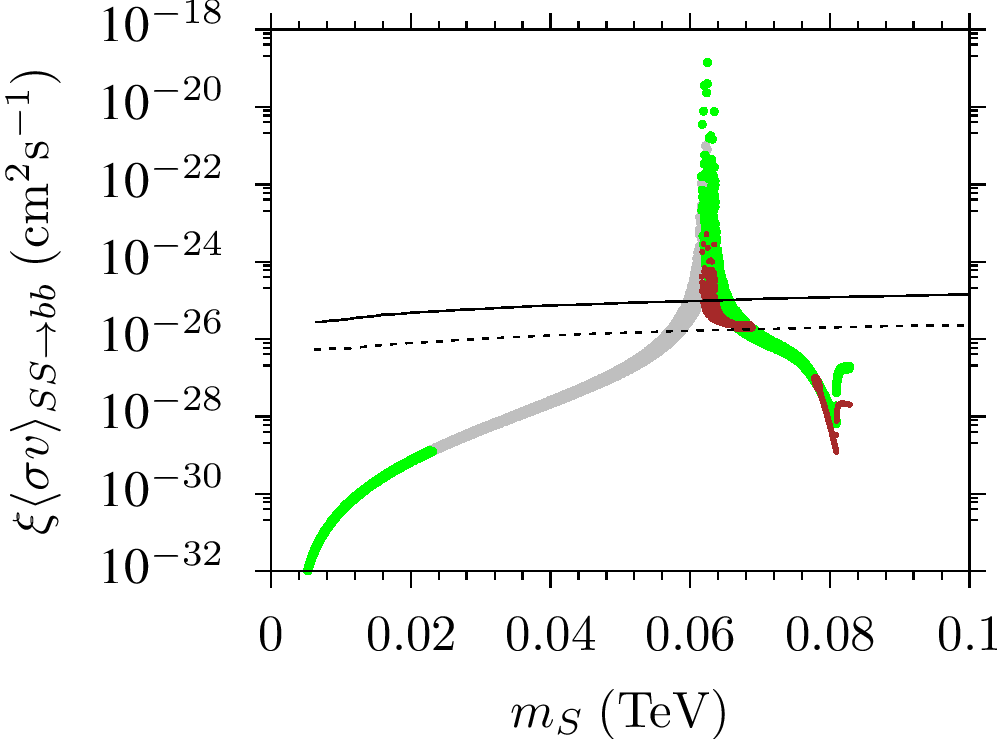}}
\caption{Average annihilation cross-section of the process $SS\rightarrow bb$ in correlation with $m_{S}$. The color coding is the same as Figure \ref{fig:SI}. The black curves represent the exclusions obtained from Fermi-LAT limits from 15 dSphs (solid), 15 years projection from 60 dSphs (dashed) \cite{Ackermann:2015zua}.}
\label{fig:sigmabb}
\end{figure}

Apart from satisfying the Planck bound on the relic abundance, some of the annihilation processes can be traced in indirect DM search experiments such as those conducted by the Fermi-LAT satellite. Figure \ref{fig:sigmabb} shows the average annihilation cross-section of the process $SS\rightarrow bb$ in correlation with $m_{S}$. The color coding is the same as Figure \ref{fig:SI}. The black curves represent the exclusions obtained from Fermi-LAT limits from 15 dSphs (solid), 15 years projection from 60 dSphs (dashed) \cite{Ackermann:2015zua}. The left panel assumes only $S$ saturates the DM relic density, and the average cross-section of annihilation into $bb$ are realized below the current exclusion bound of Fermi-Lat from 15 dSphs. The region with $60 \lesssim m_{S} \lesssim 80$ GeV lies between the current and 15 year projection exclusion bounds of Fermi-Lat, and hence, this region will be tested in near future. The right panel considers the possibility of a second DM candidate, and the brown region represents all solutions with $\Omega h^{2} \leq 0.126$. The average cross-section should be normalized with $\xi$ given in Eq.(\ref{eq:xi}). In this case, the solutions above the solid black line are already excluded by the current Fermi-LAT exclusion, while more solutions are accumulated in the region with $60 \lesssim m_{S} \lesssim 80$ GeV soon-to-be tested.

\begin{figure}[t!]
\centering
\subfigure{\includegraphics[scale=1.2]{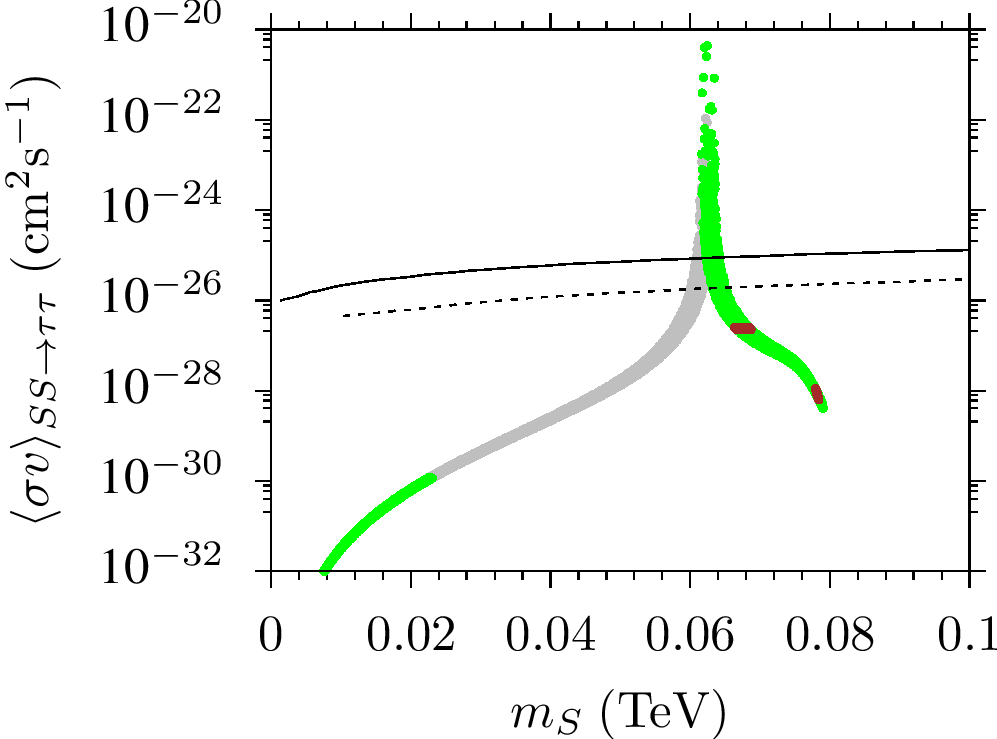}}%
\subfigure{\includegraphics[scale=1.2]{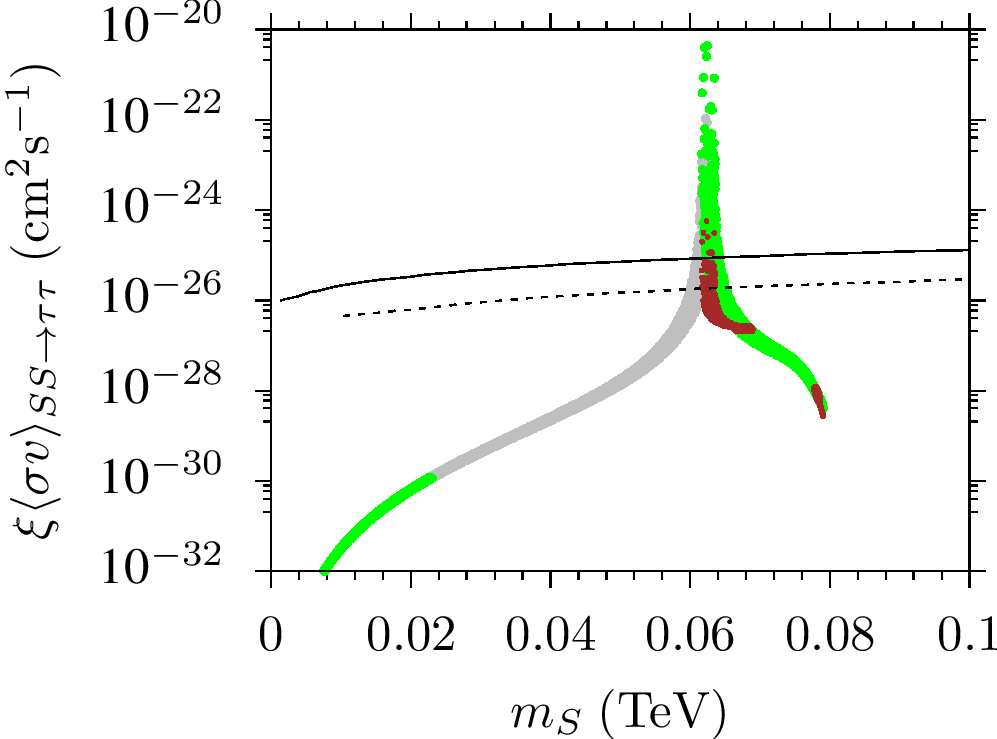}}
\caption{Average annihilation cross-section of the process $SS\rightarrow \tau\tau$ in correlation with $m_{S}$. The color coding and the curves are the same as Figure \ref{fig:sigmabb}.}
\label{fig:sigmatautau}
\end{figure}

Another annihilation channel considered in the indirect detection experiments is $SS\rightarrow\tau\tau$, which is shown in Figure \ref{fig:sigmatautau}. The color coding and the curves are the same as Figure \ref{fig:sigmabb}. When we assume $S$ to be the only DM, then its average cross-section to a pair of $\tau$ leptons lies slightly below the 15-year projection of the exclusion bound from the Fermi-LAT satellite. Such solutions can be tested in future, even though it is not as large as that of $SS\rightarrow bb$. On the other hand, if we consider another DM possibilty, then the right panel of Figure \ref{fig:sigmatautau} shows that this channel also provides testable solutions with $m_{S}\sim 60$ GeV in near future.

\begin{figure}[t!]
\centering
\subfigure{\includegraphics[scale=1.2]{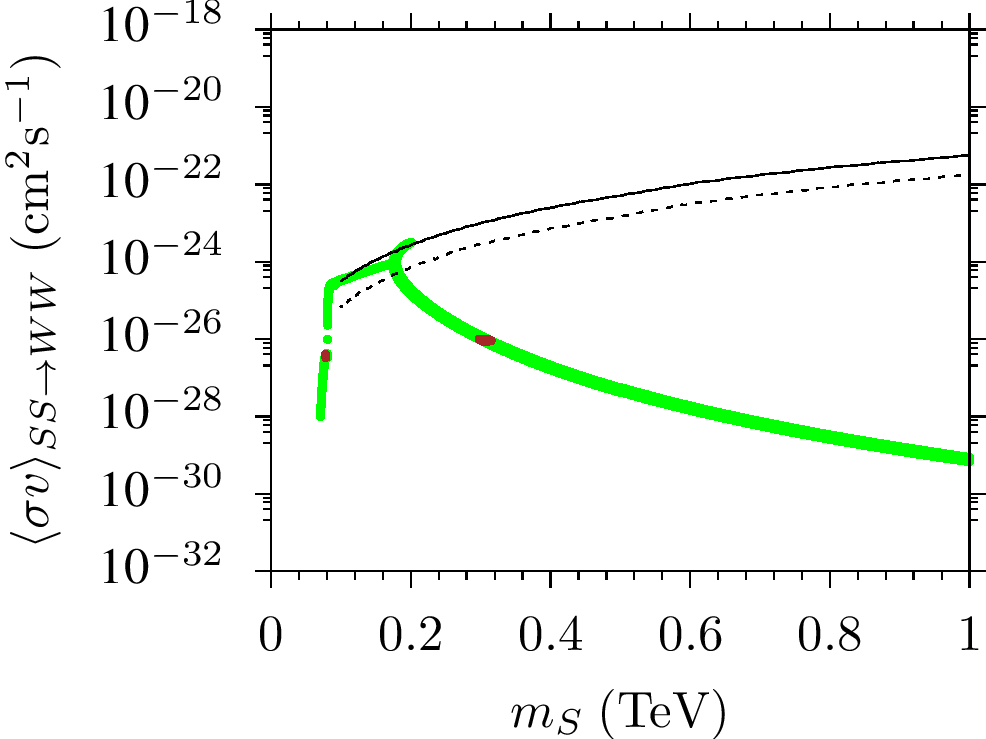}}%
\subfigure{\includegraphics[scale=1.2]{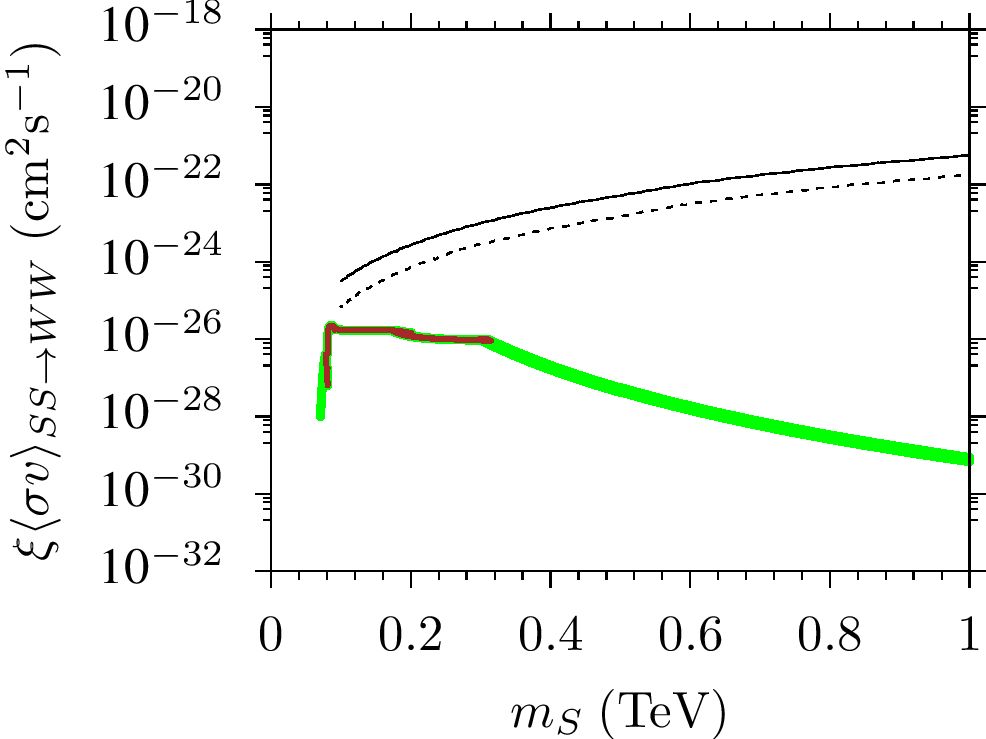}}
\caption{Average annihilation cross-section of the process $SS\rightarrow WW$ in correlation with $m_{S}$. The color coding and the curves are the same as Figure \ref{fig:sigmabb}.}
\label{fig:sigmaWW}
\end{figure}

The final channel that can be considered in the indirect DM search experiments is the one involving a pair of $W$ bosons. The average annihilation cross-section is shown in Figure \ref{fig:sigmaWW}. The color coding and the curves are the same as Figure \ref{fig:sigmabb}. Even though the exclusion bounds are sensitive to this channel, testing the model against the experimental results for $SS\rightarrow WW$ may take little longer in compared to the those involving $bb$ and $\tau\tau$, as seen from the results represented in Figure \ref{fig:sigmaWW}.

Before concluding this section, we should note that even though linkup scan does not exclude the annihilation channels mentioned above, one can obtain solutions consistent with the DM constraints in much wider range over the green regions.

%Before concluding this section we should note that the results from Figures \ref{fig:sigmabb}, \ref{fig:sigmatautau} and \ref{fig:sigmaWW} can be interpreted to the case in which $\lambda_{HS}$ is determined with Eq.(\ref{eq:fact}). Since teh data based on Eq.(\ref{eq:fact}) meet with the general analyses when $m_{S}\sim 60,80,300$ GeV, the results for $\langle \sigma v \rangle_{SS\rightarrow bb}$ and $\langle \sigma v \rangle_{SS\rightarrow \tau\tau}$ remain more or less the same, while it predicts $\langle \sigma v \rangle_{SS\rightarrow WW} \sim 10^{-27}~cm^{2}s^{-1}$.  

\section{Conclusion}
The SM, though completed experimentally, needs be extended for various reasons, ranging from dark matter to inflation. Each extension comes with its scale and mechanism, and typically pulls up the SM towards its scale at the loop level. The problem is to prevent this destabilization while allowing for the aforementioned extensions. In the case of a real singlet scalar, which may be a viable DM candidate, we discuss electroweak instability, and come to the conclusion that it can be resolved by introducing unification of quartic couplings in the potential at the point of mass denegeracy. This criterion, which we called MDDU, allows the Higgs field to couple to the singlet scalar such that heavier the scalar smaller the coupling. This seesawic coupling regime, as we have shown by simulations in Sec. 2, leads to stabilization of the Higgs boson mass. The MDDU is not a SUSY-like symmetry rule but gives a rationale for the seesawic regime needed for electroweak stability. Analysis of the DM in Sec. 3 gives a realistic view of the implications of the MDDU framework. 

From phenomenological point of view, this model can be confronted with different experimental results currently obtained from the ongoing experiments such as LHC, LUX-Zeplin and FermiLat. When the invisible Higgs boson decays is constrained to be less than $10\%$, the model predicts the Higgs boson mostly in the SM profile, while the MDDU linkup strictly forces it to be just the SM Higgs boson. The invisible decays of the Higgs boson can be constrained by externally applying the condition ${\rm BR}(h\rightarrow SS) \leq 0.10$, which requires small $\lambda_{HS}$ when $m_{S} < m_{h}$, while the linkup condition an automatically satisfy this constraint without invoking any restriction. 

If $S$ is promoted to be a candidate of DM by imposing a $Z_{2}$ symmetry, a blind scan can find solutions consistent with the current DM constraints within its whole range, while the MDDU linkup condition restricts such solutions into a very narrow range that the consistent solutions can be realized when $m_{S}=50,80$ and $300$ GeV. The first two regions correspond to approximate resonance region ($m_{S}\simeq m_{h}/2$). These solutions are expected to be tested soon in direct detection of DM experiments such as LUX-Zeplin, since they predict spin-independent DM scattering cross-section slightly below the current exclusion curve from these experiments. If one assumes only $S$ saturates the DM relic density, then the resonance solutions are excluded by the Planck bound, since $SS\rightarrow hh$ annihilation processes yield very low relic abundance for $S$. However, if the possibility of another DM sector is taken into account, then resonance solutions become available, and one can expect them to be tested soon in direct detection experiments.

We also discuss the possible annihilation channels of $S$ which take important part in satisfying the correct relic density of $S$. The relic abundance of $S$ decreases through its annihilation channels into $hh$ at about $25\%$, $tt$ at about $13\%$, $ZZ$ at about $23\%$, $WW$ at about $80\%$. These annihilation channels are realized mostly when $m_{S}$ is close by the mass of the SM particle involved. However, when $m_{S}\lesssim 80$ GeV, the $SS\rightarrow bb$ and $SS\rightarrow \tau\tau$ channels can be identified at about $80\%$ and $8\%$, respectively. These are the maximum rates of the annihilation channels, and these rate may differ at different mass scales of $S$. 

Among these annihilation channels, those involving $bb$, $\tau\tau$ and $WW$ can be traced in indirect detection experiments of DM such as FermiLat. We identify solutions for $SS\rightarrow bb$ through its average annihilation cross-section, which lie slightly below the current exclusion curve from the FermiLat experiments. Similarly, we also consider $SS\rightarrow \tau\tau$. Its results are slightly below the 15 year projection of FermiLat measurements. On the other hand, considering the possibility of another DM sector, the both annihilation channels becomes available in the current and future experiments, and even some solutions may be excluded already. We conclude the implications for FermiLat measurements with the $SS\rightarrow WW$ channel. Even though this channel is not reachable for the current experiments, it can provide testable solutions in the future upgrade of the experimental results.

The MDDU linkup (made possible by mechanisms like symmergence) can well be applied to other scalar fields like flavons, Affleck-Dine scalars, and the like. The study of the DM shows that the SM can indeed be stabilized against heavy BSM sectors provided that their scales are linked via MDDU. The electroweak stability, made possible by MDDU, allows feebly-coupled heavy BSM -- a high-luminosity challenge for collider searches. 

\section*{Acknowledgements}
The work of DD was supported in part by the T{\"U}B{\.I}TAK grant 118F387.

\label{sec:conc}


\begin{thebibliography}{99}
\bibitem{higgs}
G.~Aad {\it et al.} [ATLAS and CMS Collaborations],
  %``Measurements of the Higgs boson production and decay rates and constraints on its couplings from a combined ATLAS and CMS analysis of the LHC pp collision data at $ \sqrt{s}=7 $ and 8 TeV,''
  JHEP {\bf 1608}, 045 (2016)
  %doi:10.1007/JHEP08(2016)045
  [arXiv:1606.02266 [hep-ex]].
  %%CITATION = doi:10.1007/JHEP08(2016)045;%%
  %1142 citations counted in INSPIRE as of 20 Jan 2020
  
\bibitem{veltman}
L.~Susskind,
  %``Dynamics of Spontaneous Symmetry Breaking in the Weinberg-Salam %Theory,''
  Phys.\ Rev.\ D {\bf 20} (1979) 2619;
  %doi:10.1103/PhysRevD.20.2619
  %%CITATION = doi:10.1103/PhysRevD.20.2619;%%
  %2544 citations counted in INSPIRE as of 30 Mar 2018
M.~J.~G.~Veltman,
% {\it The Infrared - Ultraviolet Connection},
  Acta Phys.\ Polon.\ B {\bf 12} (1981) 437;
  %%CITATION = APPOA,B12,437;%%
  %673 citations counted in INSPIRE as of 21 Feb 2018
G.~F.~Giudice,
  %``Naturalness after LHC8,''
  PoS EPS  (2013) 163
  [arXiv:1307.7879 [hep-ph]].
  %%CITATION = ARXIV:1307.7879;%%
  %80 citations counted in INSPIRE as of 30 Mar 2018

\bibitem{grav}
G.~'t Hooft,
%  {\it Obstacles on the way towards the Quantization of Space, Time And Matter - and possible resolutions},
  Stud.\ Hist.\ Phil.\ Sci.\ B {\bf 32} (2001) 157;
  %doi:10.1016/S1355-2198(01)00008-9
  %%CITATION = doi:10.1016/S1355-2198(01)00008-9;%%
  %7 citations counted in INSPIRE as of 21 Feb 2018
R.~M.~Wald,
  %``The Formulation of Quantum Field Theory in Curved Spacetime,''
  Einstein Stud.\  {\bf 14}, 439 (2018)
  %doi:10.1007/978-1-4939-7708-6_15
  [arXiv:0907.0416 [gr-qc]].
  %%CITATION = doi:10.1007/978-1-4939-7708-6_15;%%
  %13 citations counted in INSPIRE as of 09 Jan 2020  
  
\bibitem{dm-ilk}
V.~C.~Rubin, W.~K.~Ford,~Jr., N.~Thonnard, and M.~Roberts,
%{\it Motion of the Galaxy and the Local Group Determined From the Velocity Anisotropy of Distant SC I Galaxies I}, 
Astron.\ J.\ {\bf 81} (1976) 687.

\bibitem{exotica} 
S.~Rappoccio,
%``The experimental status of direct searches for exotic physics beyond the standard model at the Large Hadron Collider,''
Rev. Phys. \textbf{4}, 100027 (2019)
%doi:10.1016/j.revip.2018.100027
[arXiv:1810.10579 [hep-ex]].
%7 citations counted in INSPIRE as of 19 Apr 2020


\bibitem{demir1}
D.~A.~Demir,
  %``Curvature-Restored Gauge Invariance and Ultraviolet Naturalness,''
  Adv.\ High Energy Phys.\  {\bf 2016}, 6727805 (2016)
  %doi:10.1155/2016/6727805
  [arXiv:1605.00377 [hep-ph]];
  %%CITATION = doi:10.1155/2016/6727805;%%
  %6 citations counted in INSPIRE as of 28 Jan 2020  
\bibitem{demir2}
D.~Demir,  
  %``Symmergent Gravity, Seesawic New Physics, and their Experimental Signatures,''
  Adv.\ High Energy Phys.\  {\bf 2019}, 4652048 (2019)
  %doi:10.1155/2019/4652048
  [arXiv:1901.07244 [hep-ph]].
  %%CITATION = doi:10.1155/2019/4652048;%%
  %2 citations counted in INSPIRE as of 28 Jan 2020


\bibitem{dm-review}  
J.~Liu, X.~Chen and X.~Ji,
  %``Current status of direct dark matter detection experiments,''
  Nature Phys.\  {\bf 13} (2017) 212
  %doi:10.1038/nphys4039
  [arXiv:1709.00688 [astro-ph.CO]];
  %%CITATION = doi:10.1038/nphys4039;%%
  %22 citations counted in INSPIRE as of 30 Mar 2018
L.~Baudis,
  %``The Search for Dark Matter,''
  European Review\  {\bf 26} (2018) 70
  %doi:10.1017/S1062798717000783
  [arXiv:1801.08128 [astro-ph.CO]].
  %%CITATION = doi:10.1017/S1062798717000783;%%
  %1 citations counted in INSPIRE as of 29 Aug 2018


\bibitem{dm00}
V.~Silveira and A.~Zee,
%``SCALAR PHANTOMS,''
Phys.\ Lett.\ B \textbf{161}, 136-140 (1985)
%doi:10.1016/0370-2693(85)90624-0
%607 citations counted in INSPIRE as of 01 Apr 2020

\bibitem{dm01}
J.~McDonald,
  %``Gauge singlet scalars as cold dark matter,''
  Phys.\ Rev.\ D {\bf 50}, 3637 (1994)
  %doi:10.1103/PhysRevD.50.3637
  [hep-ph/0702143 [HEP-PH]].
  %%CITATION = doi:10.1103/PhysRevD.50.3637;%%
  %741 citations counted in INSPIRE as of 01 Apr 2020

\bibitem{demir0}
D.~A.~Demir,
%``Weak scale hidden sector and electroweak Q balls,''
Phys.\ Lett.\ B \textbf{450}, 215-219 (1999)
%doi:10.1016/S0370-2693(99)00106-9
[arXiv:hep-ph/9810453 [hep-ph]].
%25 citations counted in INSPIRE as of 01 Apr 2020

\bibitem{dm0}
C.~Burgess, M.~Pospelov and T.~ter Veldhuis,
%``The Minimal model of nonbaryonic dark matter: A Singlet scalar,''
Nucl.\ Phys.\ B \textbf{619}, 709-728 (2001)
%doi:10.1016/S0550-3213(01)00513-2
[arXiv:hep-ph/0011335 [hep-ph]].
%799 citations counted in INSPIRE as of 01 Apr 2020


\bibitem{no-S} 
B.~Vachon [ATLAS and CMS Collaborations],
  %``Exploring the Standard Model at the LHC,''
  Int.\ J.\ Mod.\ Phys.\ A {\bf 31}, 1630034 (2016).
  %doi:10.1142/S0217751X16300349
  %%CITATION = doi:10.1142/S0217751X16300349;%%

\bibitem{g-m}
G.~Giudice and A.~Masiero,
%``A Natural Solution to the mu Problem in Supergravity Theories,''
Phys. Lett. B \textbf{206}, 480-484 (1988)
%doi:10.1016/0370-2693(88)91613-9
%969 citations counted in INSPIRE as of 27 Apr 2020

\bibitem{demir3}  
D.~Demir,
  {\it Naturalizing Gravity of the Quantum Fields, and the Hierarchy Problem},
  arXiv:1703.05733 [hep-ph].
  %%CITATION = ARXIV:1703.05733;%%
  %4 citations counted in INSPIRE as of 28 Jan 2020


\bibitem{cankocak}
K.~Cankoçak, D.~Demir, C.~Karahan and S.~Şen,
{\it Electroweak Stability and Discovery Luminosities for New Physics},
[arXiv:2002.12262 [hep-ph]].
%0 citations counted in INSPIRE as of 02 Apr 2020


%\cite{Haba:2016gqx}
\bibitem{Haba:2016gqx} 
  N.~Haba, H.~Ishida, N.~Okada and Y.~Yamaguchi,
  %``Multiple-point principle with a scalar singlet extension of the Standard Model,''
  PTEP {\bf 2017}, no. 1, 013B03 (2017)
  doi:10.1093/ptep/ptw186
  [arXiv:1608.00087 [hep-ph]], and references therein.
  %%CITATION = doi:10.1093/ptep/ptw186;%%
  %14 citations counted in INSPIRE as of 29 Feb 2020

\bibitem{meta}
J.~Elias-Miro, J.~R.~Espinosa, G.~F.~Giudice, G.~Isidori, A.~Riotto and A.~Strumia,
%``Higgs mass implications on the stability of the electroweak vacuum,''
Phys.\ Lett.\ B \textbf{709}, 222-228 (2012)
%doi:10.1016/j.physletb.2012.02.013
[arXiv:1112.3022 [hep-ph]];
%454 citations counted in INSPIRE as of 03 Apr 2020
G.~Degrassi, S.~Di Vita, J.~Elias-Miro, J.~R.~Espinosa, G.~F.~Giudice, G.~Isidori and A.~Strumia,
%``Higgs mass and vacuum stability in the Standard Model at NNLO,''
JHEP \textbf{08}, 098 (2012)
%doi:10.1007/JHEP08(2012)098
[arXiv:1205.6497 [hep-ph]].
%1273 citations counted in INSPIRE as of 03 Apr 2020

\bibitem{Planck}
N.~Aghanim \textit{et al.} [Planck],
%``Planck 2018 results. VI. Cosmological parameters,''
[arXiv:1807.06209 [astro-ph.CO]].
%2590 citations counted in INSPIRE as of 25 Apr 2020

%\cite{PerezLorenzana:1998rj}
\bibitem{PerezLorenzana:1998rj} See, for instance, \\
  A.~Perez-Lorenzana, W.~A.~Ponce and A.~Zepeda,
  %``NonSUSY unification in left-right models,''
  Phys.\ Rev.\ D {\bf 59}, 116004 (1999)
  doi:10.1103/PhysRevD.59.116004
  [hep-ph/9812401];
  %%CITATION = doi:10.1103/PhysRevD.59.116004;%%
  %8 citations counted in INSPIRE as of 27 Mar 2020
%\cite{Shaban:1992he}
%\bibitem{Shaban:1992he} 
  N.~T.~Shaban and W.~J.~Stirling,
  %``Minimal left-right symmetry and SO(10) grand unification using LEP coupling constant measurements,''
  Phys.\ Lett.\ B {\bf 291}, 281 (1992).
  doi:10.1016/0370-2693(92)91046-C;
  %%CITATION = doi:10.1016/0370-2693(92)91046-C;%%
  %27 citations counted in INSPIRE as of 27 Mar 2020
%\cite{LalAwasthi:2011aa}
%\bibitem{LalAwasthi:2011aa} 
  R.~Lal Awasthi and M.~K.~Parida,
  %``Inverse Seesaw Mechanism in Nonsupersymmetric SO(10), Proton Lifetime, Nonunitarity Effects, and a Low-mass Z' Boson,''
  Phys.\ Rev.\ D {\bf 86}, 093004 (2012)
  doi:10.1103/PhysRevD.86.093004
  [arXiv:1112.1826 [hep-ph]].
  %%CITATION = doi:10.1103/PhysRevD.86.093004;%%
  %36 citations counted in INSPIRE as of 27 Mar 2020




%\cite{Martin:2003qz}
\bibitem{Martin:2003qz} 
  S.~P.~Martin,
  %``Evaluation of two loop selfenergy basis integrals using differential equations,''
  Phys.\ Rev.\ D {\bf 68}, 075002 (2003)
  doi:10.1103/PhysRevD.68.075002
  [hep-ph/0307101];
  %%CITATION = doi:10.1103/PhysRevD.68.075002;%%
  %126 citations counted in INSPIRE as of 02 Mar 2020
%\cite{Martin:2003it}
%\bibitem{Martin:2003it} 
  S.~P.~Martin,
  %``Two loop scalar self energies in a general renormalizable theory at leading order in gauge couplings,''
  Phys.\ Rev.\ D {\bf 70}, 016005 (2004)
  doi:10.1103/PhysRevD.70.016005
  [hep-ph/0312092].
  %%CITATION = doi:10.1103/PhysRevD.70.016005;%%
  %114 citations counted in INSPIRE as of 02 Mar 2020

%\cite{Ferreira:2015pfi}
\bibitem{Ferreira:2015pfi} 
  P.~M.~Ferreira and B.~Swiezewska,
  %``One-loop contributions to neutral minima in the inert doublet model,''
  JHEP {\bf 1604}, 099 (2016)
  doi:10.1007/JHEP04(2016)099
  [arXiv:1511.02879 [hep-ph]].
  %%CITATION = doi:10.1007/JHEP04(2016)099;%%
  %24 citations counted in INSPIRE as of 02 Mar 2020

%\cite{Degrassi:2002fi}
\bibitem{Degrassi:2002fi}
G.~Degrassi, S.~Heinemeyer, W.~Hollik, P.~Slavich and G.~Weiglein,
%``Towards high precision predictions for the MSSM Higgs sector,''
Eur. Phys. J. C \textbf{28}, 133-143 (2003)
doi:10.1140/epjc/s2003-01152-2
[arXiv:hep-ph/0212020 [hep-ph]].
%1046 citations counted in INSPIRE as of 25 Apr 2020


%\cite{Group:2009ad}
\bibitem{Group:2009ad} 
  Tevatron Electroweak Working Group [CDF and D0 Collaborations],
  %``Combination of CDF and D0 Results on the Mass of the Top Quark,''
  arXiv:0903.2503 [hep-ex].
  %%CITATION = ARXIV:0903.2503;%%
  %325 citations counted in INSPIRE as of 04 Mar 2020

%\cite{Gogoladze:2011aa}
\bibitem{Gogoladze:2011aa} 
  I.~Gogoladze, Q.~Shafi and C.~S.~Un,
  %``Higgs Boson Mass from t-b-$\tau$ Yukawa Unification,''
  JHEP {\bf 1208}, 028 (2012)
  doi:10.1007/JHEP08(2012)028
  [arXiv:1112.2206 [hep-ph]];
  %%CITATION = doi:10.1007/JHEP08(2012)028;%%
  %69 citations counted in INSPIRE as of 04 Mar 2020
%\cite{Ajaib:2013zha}
%\bibitem{Ajaib:2013zha} 
  M.~Adeel Ajaib, I.~Gogoladze, Q.~Shafi and C.~S.~Un,
  %``A Predictive Yukawa Unified SO(10) Model: Higgs and Sparticle Masses,''
  JHEP {\bf 1307}, 139 (2013)
  doi:10.1007/JHEP07(2013)139
  [arXiv:1303.6964 [hep-ph]].
  %%CITATION = doi:10.1007/JHEP07(2013)139;%%
  %49 citations counted in INSPIRE as of 04 Mar 2020

%\cite{Porod:2003um}
\bibitem{Porod:2003um}
  W.~Porod,
  %``SPheno, a program for calculating supersymmetric spectra, SUSY particle decays and SUSY particle production at e+ e- colliders,''
  Comput.\ Phys.\ Commun.\  {\bf 153}, 275 (2003)
  doi:10.1016/S0010-4655(03)00222-4
  [hep-ph/0301101];
  %%CITATION = doi:10.1016/S0010-4655(03)00222-4;%%
  %944 citations counted in INSPIRE as of 27 Mar 2020
%\cite{Porod:2011nf}
%\bibitem{Porod:2011nf}
  W.~Porod and F.~Staub,
  %``SPheno 3.1: Extensions including flavour, CP-phases and models beyond the MSSM,''
  Comput.\ Phys.\ Commun.\  {\bf 183}, 2458 (2012)
  doi:10.1016/j.cpc.2012.05.021
  [arXiv:1104.1573 [hep-ph]];
  %%CITATION = doi:10.1016/j.cpc.2012.05.021;%%
  %585 citations counted in INSPIRE as of 27 Mar 2020


%\cite{Staub:2013tta}
\bibitem{Staub:2013tta} 
  F.~Staub,
  %``SARAH 4 : A tool for (not only SUSY) model builders,''
  Comput.\ Phys.\ Commun.\  {\bf 185}, 1773 (2014)
  doi:10.1016/j.cpc.2014.02.018
  [arXiv:1309.7223 [hep-ph]];
  %%CITATION = doi:10.1016/j.cpc.2014.02.018;%%
  %496 citations counted in INSPIRE as of 27 Mar 2020
%\cite{Staub:2015kfa}
%\bibitem{Staub:2015kfa} 
  F.~Staub,
  %``Exploring new models in all detail with SARAH,''
  Adv.\ High Energy Phys.\  {\bf 2015}, 840780 (2015)
  doi:10.1155/2015/840780
  [arXiv:1503.04200 [hep-ph]].
  %%CITATION = doi:10.1155/2015/840780;%%
  %130 citations counted in INSPIRE as of 27 Mar 2020

%\cite{Belanger:2001fz}
\bibitem{Belanger:2001fz} 
  G.~Belanger, F.~Boudjema, A.~Pukhov and A.~Semenov,
  %``MicrOMEGAs: A Program for calculating the relic density in the MSSM,''
  Comput.\ Phys.\ Commun.\  {\bf 149}, 103 (2002)
  doi:10.1016/S0010-4655(02)00596-9
  [hep-ph/0112278];
  %%CITATION = doi:10.1016/S0010-4655(02)00596-9;%%
  %597 citations counted in INSPIRE as of 27 Mar 2020
%\cite{Belanger:2006is}
%\bibitem{Belanger:2006is} 
  G.~Belanger, F.~Boudjema, A.~Pukhov and A.~Semenov,
  %``MicrOMEGAs 2.0: A Program to calculate the relic density of dark matter in a generic model,''
  Comput.\ Phys.\ Commun.\  {\bf 176}, 367 (2007)
  doi:10.1016/j.cpc.2006.11.008
  [hep-ph/0607059];
  %%CITATION = doi:10.1016/j.cpc.2006.11.008;%%
  %651 citations counted in INSPIRE as of 27 Mar 2020
%\cite{Belanger:2013oya}
%\bibitem{Belanger:2013oya} 
  G.~Belanger, F.~Boudjema, A.~Pukhov and A.~Semenov,
  %``micrOMEGAs_3: A program for calculating dark matter observables,''
  Comput.\ Phys.\ Commun.\  {\bf 185}, 960 (2014)
  doi:10.1016/j.cpc.2013.10.016
  [arXiv:1305.0237 [hep-ph]].
  %%CITATION = doi:10.1016/j.cpc.2013.10.016;%%
  %586 citations counted in INSPIRE as of 27 Mar 2020

%\cite{Akrami:2018vks}
\bibitem{Akrami:2018vks} 
  Y.~Akrami {\it et al.} [Planck Collaboration],
  %``Planck 2018 results. I. Overview and the cosmological legacy of Planck,''
  arXiv:1807.06205 [astro-ph.CO].
  %%CITATION = ARXIV:1807.06205;%%
  %294 citations counted in INSPIRE as of 04 Mar 2020

%\cite{CMS:2019bke}
\bibitem{CMS:2019bke} 
  CMS Collaboration [CMS Collaboration],
  %``First constraints on invisible Higgs boson decays using $\mathrm{t}\bar{\mathrm{t}}\mathrm{H}$ production at $\sqrt{s}=13~\mathrm{TeV}$,''
  CMS-PAS-HIG-18-008.
  %%CITATION = CMS-PAS-HIG-18-008;%%
  %5 citations counted in INSPIRE as of 27 Mar 2020

%\cite{Aaboud:2018sfi}
\bibitem{Aaboud:2018sfi} 
  M.~Aaboud {\it et al.} [ATLAS Collaboration],
  %``Search for invisible Higgs boson decays in vector boson fusion at $\sqrt{s} = 13$ TeV with the ATLAS detector,''
  Phys.\ Lett.\ B {\bf 793}, 499 (2019)
  doi:10.1016/j.physletb.2019.04.024
  [arXiv:1809.06682 [hep-ex]].
  %%CITATION = doi:10.1016/j.physletb.2019.04.024;%%
  %40 citations counted in INSPIRE as of 27 Mar 2020



%\cite{Sirunyan:2018owy}
\bibitem{Sirunyan:2018owy} 
  A.~M.~Sirunyan {\it et al.} [CMS Collaboration],
  %``Search for invisible decays of a Higgs boson produced through vector boson fusion in proton-proton collisions at $\sqrt{s} =$ 13 TeV,''
  Phys.\ Lett.\ B {\bf 793}, 520 (2019)
  doi:10.1016/j.physletb.2019.04.025
  [arXiv:1809.05937 [hep-ex]];
  %%CITATION = doi:10.1016/j.physletb.2019.04.025;%%
  %96 citations counted in INSPIRE as of 27 Mar 2020
%\cite{Aad:2015pla}
%\bibitem{Aad:2015pla} 
  G.~Aad {\it et al.} [ATLAS Collaboration],
  %``Constraints on new phenomena via Higgs boson couplings and invisible decays with the ATLAS detector,''
  JHEP {\bf 1511}, 206 (2015)
  doi:10.1007/JHEP11(2015)206
  [arXiv:1509.00672 [hep-ex]];
  %%CITATION = doi:10.1007/JHEP11(2015)206;%%
  %303 citations counted in INSPIRE as of 27 Mar 2020
%\cite{Chatrchyan:2008aa}
%\bibitem{Chatrchyan:2008aa} 
  S.~Chatrchyan {\it et al.} [CMS Collaboration],
  %``The CMS Experiment at the CERN LHC,''
  JINST {\bf 3}, S08004 (2008).
  doi:10.1088/1748-0221/3/08/S08004;
  %%CITATION = doi:10.1088/1748-0221/3/08/S08004;%%
  %6807 citations counted in INSPIRE as of 27 Mar 2020
%\cite{Khachatryan:2016whc}
%\bibitem{Khachatryan:2016whc} 
  V.~Khachatryan {\it et al.} [CMS Collaboration],
  %``Searches for invisible decays of the Higgs boson in pp collisions at $\sqrt{s}$ = 7, 8, and 13 TeV,''
  JHEP {\bf 1702}, 135 (2017)
  doi:10.1007/JHEP02(2017)135
  [arXiv:1610.09218 [hep-ex]].
  %%CITATION = doi:10.1007/JHEP02(2017)135;%%
  %194 citations counted in INSPIRE as of 27 Mar 2020


%\cite{Djouadi:2011aa}
\bibitem{Djouadi:2011aa} 
  A.~Djouadi, O.~Lebedev, Y.~Mambrini and J.~Quevillon,
  %``Implications of LHC searches for Higgs--portal dark matter,''
  Phys.\ Lett.\ B {\bf 709}, 65 (2012)
  doi:10.1016/j.physletb.2012.01.062
  [arXiv:1112.3299 [hep-ph]].
  %%CITATION = doi:10.1016/j.physletb.2012.01.062;%%
  %408 citations counted in INSPIRE as of 27 Mar 2020

%\cite{Belanger:2013xza}
\bibitem{Belanger:2013xza} 
  G.~Belanger, B.~Dumont, U.~Ellwanger, J.~F.~Gunion and S.~Kraml,
  %``Global fit to Higgs signal strengths and couplings and implications for extended Higgs sectors,''
  Phys.\ Rev.\ D {\bf 88}, 075008 (2013)
  doi:10.1103/PhysRevD.88.075008
  [arXiv:1306.2941 [hep-ph]].
  %%CITATION = doi:10.1103/PhysRevD.88.075008;%%
  %280 citations counted in INSPIRE as of 27 Mar 2020



%\cite{Giacchino:2013bta}
%\bibitem{Giacchino:2013bta} 
%  F.~Giacchino, L.~Lopez-Honorez and M.~H.~G.~Tytgat,
  %``Scalar Dark Matter Models with Significant Internal Bremsstrahlung,''
%  JCAP {\bf 1310}, 025 (2013)
%  doi:10.1088/1475-7516/2013/10/025
%  [arXiv:1307.6480 [hep-ph]].
  %%CITATION = doi:10.1088/1475-7516/2013/10/025;%%
  %63 citations counted in INSPIRE as of 07 Mar 2020

%\cite{Brink:2005ej}
\bibitem{Brink:2005ej} 
  P.~L.~Brink {\it et al.} [CDMS-II Collaboration],
  %``Beyond the CDMS-II dark matter search: SuperCDMS,''
  eConf C {\bf 041213}, 2529 (2004)
  [astro-ph/0503583].
  %%CITATION = ASTRO-PH/0503583;%%
  %80 citations counted in INSPIRE as of 07 Mar 2020

%\cite{Akerib:2018dfk}
\bibitem{Akerib:2018dfk} 
  D.~S.~Akerib {\it et al.} [LUX-ZEPLIN Collaboration],
  %``Projected WIMP Sensitivity of the LUX-ZEPLIN (LZ) Dark Matter Experiment,''
  Phys.\ Rev.\ D {\bf 101}, 052002 (2020)
  doi:10.1103/PhysRevD.101.052002
  [arXiv:1802.06039 [astro-ph.IM]].
  %%CITATION = doi:10.1103/PhysRevD.101.052002;%%
  %124 citations counted in INSPIRE as of 07 Mar 2020

%\cite{Belanger:2015vwa}
\bibitem{Belanger:2015vwa} 
  G.~Belanger, D.~Ghosh, R.~Godbole and S.~Kulkarni,
  %``Light stop in the MSSM after LHC Run 1,''
  JHEP {\bf 1509}, 214 (2015)
  doi:10.1007/JHEP09(2015)214
  [arXiv:1506.00665 [hep-ph]].
  %%CITATION = doi:10.1007/JHEP09(2015)214;%%
  %35 citations counted in INSPIRE as of 07 Mar 2020

%\cite{Ackermann:2015zua}
\bibitem{Ackermann:2015zua} 
  M.~Ackermann {\it et al.} [Fermi-LAT Collaboration],
  %``Searching for Dark Matter Annihilation from Milky Way Dwarf Spheroidal Galaxies with Six Years of Fermi Large Area Telescope Data,''
  Phys.\ Rev.\ Lett.\  {\bf 115}, no. 23, 231301 (2015)
  doi:10.1103/PhysRevLett.115.231301
  [arXiv:1503.02641 [astro-ph.HE]];
  %%CITATION = doi:10.1103/PhysRevLett.115.231301;%%
  %864 citations counted in INSPIRE as of 07 Mar 2020
%\cite{Drlica-Wagner:2015xua}
%\bibitem{Drlica-Wagner:2015xua} 
  A.~Drlica-Wagner {\it et al.} [Fermi-LAT and DES Collaborations],
  %``Search for Gamma-Ray Emission from DES Dwarf Spheroidal Galaxy Candidates with Fermi-LAT Data,''
  Astrophys.\ J.\  {\bf 809}, no. 1, L4 (2015)
  doi:10.1088/2041-8205/809/1/L4
  [arXiv:1503.02632 [astro-ph.HE]].
  %%CITATION = doi:10.1088/2041-8205/809/1/L4;%%
  %154 citations counted in INSPIRE as of 07 Mar 2020

%\cite{Aartsen:2017ulx}
%\bibitem{Aartsen:2017ulx} 
%  M.~G.~Aartsen {\it et al.} [IceCube Collaboration],
  %``Search for Neutrinos from Dark Matter Self-Annihilations in the center of the Milky Way with 3 years of IceCube/DeepCore,''
%  Eur.\ Phys.\ J.\ C {\bf 77}, no. 9, 627 (2017)
%  doi:10.1140/epjc/s10052-017-5213-y
%  [arXiv:1705.08103 [hep-ex]]; \\
  %%CITATION = doi:10.1140/epjc/s10052-017-5213-y;%%
  %68 citations counted in INSPIRE as of 07 Mar 2020
%also see the talk by Sebastian Baur in NDM 2020, \\
%\href{https://indico.cern.ch/event/813648/contributions/3666175/attachments/1969561/3275879/Sun12_Exp_sebastian-baur.pdf}{Indirect searches for Dark Matter with the IceCube neutrino telescope}.





\end{thebibliography}
\end{document}